\documentclass[a4,referee,pre,showpacs]{revtex4}

\def\EQ{\begin{equation}}
\def\EN{\end{equation}}
\def\EQA{\begin{eqnarray}}
\def\ENA{\end{eqnarray}}

\usepackage{amsmath}
\usepackage[dvips]{graphicx}

\def\ie{{\em i.e. }}

\begin{document}

\title{Dynamics and thermodynamics of axisymmetric flows:\\
I. Theory}
\author{N. Leprovost \footnote{present address: Laboratoire de physique statistique de l'ENS, \\ (UMR 8550) 24 rue Lhomond, F-75231 Paris cedex 05} and B. Dubrulle}
\affiliation{DRECAM/SPEC/CEA Saclay, and CNRS (URA2464), F-91190 Gif
sur Yvette Cedex, France}
\author{P.-H. Chavanis}
\affiliation{Laboratoire de Physique Th\'eorique (UMR 5152),
Universit\'e Paul Sabatier, 118, route de Narbonne
31062 Toulouse, France}

\begin{abstract}

We develop new variational principles to study the structure and the
stability of equilibrium states of axisymmetric flows. We show that
the axisymmetric Euler equations for inviscid flows admit an infinite
number of steady state solutions. We find their general form and
provide analytical solutions in some special cases. The system can be
trapped in one of these steady states as a result of an inviscid
violent relaxation.  We show that the stable steady states maximize a
(non-universal) $H$-function while conserving energy, helicity,
circulation and angular momentum (robust constraints). This can be
viewed as a form of generalized selective decay principle.  We derive
relaxation equations which can be used as numerical algorithm to
construct nonlinearly dynamically stable stationary solutions of
axisymmetric flows. We also develop a thermodynamical approach to
predict the equilibrium state at some fixed coarse-grained scale. We
show that the resulting distribution can be divided in two parts: one
universal coming from the conservation of robust invariants and one
non-universal determined by the initial conditions through the fragile
invariants (for freely evolving systems) or by a prior distribution
encoding non-ideal effects such as viscosity, small-scale forcing and
dissipation (for forced systems). Finally, we derive a
parameterization of inviscid mixing to describe the dynamics of the
system at the coarse-grained scale. A conceptual interest of this
axisymmetric model is to be intermediate between 2D and 3D turbulence.
\vspace{1cm}

\end{abstract}


\vspace{0.1cm}

\maketitle

\section{Introduction}

The ubiquity of rotating systems in astrophysics and geophysics makes
axisymmetric flows a classical paradigm. In the laboratory, two
axisymmetric devices, the Taylor-Couette flow and the von K\'arm\'an
flow, have become a standard to investigate issues such as
super-critical and sub-critical stability \cite{Prigent02},
fluctuation of global quantities \cite{Aumaitre01,Labbe96b} or
turbulent transport \cite{Taylor23,Lathrop92,Marie04}. However, many
basic issues regarding stability and turbulence in axisymmetric
flows still remain unsolved. For example, one still fails to
understand the onset of turbulence or the equilibrium state  in
Taylor-Couette with outer rotating cylinder \cite{Dubrulle04}, or the
recent bifurcation of the turbulent state observed in von K\'arm\'an flow
\cite{Ravelet04}.\

In the past, dynamical stability and equilibrium properties of flows have often
been studied using variational \cite{Chandrasekhar61} or maximization
\cite{Busse96} principles. Examples of application to axisymmetric
flows include necessary criteria for instability or turbulent velocity
profiles in Taylor-Couette flow. Maximization or minimization
principles have also been used to give sufficient criteria of {\it
nonlinear dynamical stability} \cite{Holm85}. One interest of these
methods is their robustness, in the sense that they mostly depend on
characteristic global quantities of the system (such as the energy)
but not necessarily on small-scale dissipation or boundary
conditions. More recently, optimization methods have been developed
within the framework of statistical mechanics for two-dimensional (2D)
perfect fluids. In that case, variational principles based on entropy
maximization determine conditions of {\it thermodynamical
stability}. Onsager
\cite{Onsager49} first used a Hamiltonian model of point vortices and
identified turbulence as a state of negative temperature leading to
the coalescence of vortices of same sign \cite{Montgomery74}. Further
improvements were provided by Kuzmin \cite{Kuzmin82}, Miller
\cite{Miller90} and Robert and Sommeria \cite{Robert91} who
independently introduced a discretization of the vorticity in a
certain number of levels to account for the continuous nature of
vorticity. Using the maximum entropy formalism of statistical
mechanics \cite{Jaynes57}, it is then possible to obtain the shape of
the metaequilibrium solution of Euler's equation as well as the
distribution of the fine-grained fluctuations around it. A variety of
solutions are found and the bifurcation diagram displays a rich
structure as illustrated by Chavanis \& Sommeria \cite{JFM1} in a
particular limit of the statistical theory. Two-dimensional turbulence
is however very peculiar since it misses vortex stretching, one
essential ingredient in 3D turbulence. The adaptation of these methods
to more realistic situations is therefore not obvious.

In the case where the system admits a scalar invariant ($ D_t
\sigma = \partial_t \sigma + {\bf u} \cdot {\bf \nabla} \sigma = 0$),
one can show that a Liouville theorem holds (incompressibility of the
motion in phase space). Indeed, the proof given by Kraichnan \&
Montgomery \cite{Kraichnan80} in the case of 2D turbulence can be
extended to any dimensional turbulence with a conserved quantity. This
is in fact the case for axisymmetric flows where the symmetry imposes
angular momentum conservation. Due to violent relaxation, the system
is expected to reach a metaequilibrium state which is a steady solution of the
axisymmetric Euler equations.  The purpose
of the present paper is to derive explicit results regarding
nonlinear dynamical stability and thermodynamical equilibrium
properties of axisymmetric flows using optimization methods.

In the first part of the paper (Sec. \ref{SectionDyn}), we consider
the nonlinear dynamical stability of stationary solutions of the
axisymmetric Euler equations (\ie without viscosity). These equations
are written in Sec. \ref{SectionAxi} and the general form of
stationary solutions is obtained in Sec. \ref{SectionStationary}. In
Sec. \ref{SectionConservation}, we list the conservation laws of the
axisymmetric Euler equations. We find non trivial invariants in
addition to the usual ones. In Sec. \ref{SectionDynstab}, we show that
the equilibrium solutions can be obtained by extremizing a functional
built with all the invariants. The fact that this optimization
procedure returns the general stationary solution means that we have
found all the invariants. In Sec. \ref{SectionSelective}, we
distinguish between fragile (Casimirs) and robust (energy, helicity,
circulation, angular momentum...) invariants. We argue that, in the
presence of viscosity or coarse-graining, the metaequilibrium state
maximizes a certain (non-universal) H-function while conserving the
robust constraints. This is similar to the case of pure 2D
hydrodynamics \cite{Chavanis03,prior,kupka,super} except for the replacement
of vorticity by angular momentum. In Sec. \ref{SectionAlgo}, we
propose a numerical algorithm based on the maximization of the
production of an H-function while conserving the robust
constraints. This can be used to compute numerically arbitrary
nonlinearly dynamically stable stationary solutions of the
axisymmetric Euler equations. This is similar to the relaxation
equations proposed by Chavanis \cite{Chavanis03,prior,kupka} in pure
2D hydrodynamics. In Sec. \ref{SectionSolution}, we provide simple
analytical steady solutions of axisymmetric equilibrium flows. In the
second part of the paper (Sec. \ref{SectionStat}), we develop the
statistical mechanics of such flows to predict the metaequilibrium
state. The statistical equilibrium state is obtained by maximizing a
mixing entropy (Sec. \ref{SectionEntropy}) while taking into account
all the constraints of the dynamics. This yields a Gibbs state
(Sec. \ref{SectionGibbs}) which gives the equilibrium coarse-grained
angular momentum as well as the fluctuations around it.  We check that
the coarse-grained field is a stationary solution of the axisymmetric
Euler equations. However, since the Casimirs are not conserved on the
coarse-grained scale, the distribution of fluctuations is non
universal and depends on the initial conditions (or fine-grained
constraints). This is also the case for the coarse-grained field. In
Sec. \ref{SectionMEPP}, we use a maximization of the entropy
production to derive relaxation equations towards the statistical
equilibrium state. This is similar to the approach proposed by Robert
\& Sommeria \cite{Robert92} in pure 2D hydrodynamics. Finally, in
Sec. \ref{SectionPrior}, we introduce the notion of prior distribution
of fluctuations for systems that are forced at small scales.  We show
that the coarse-grained field maximizes at statistical equilibrium a
generalized entropy fixed by the prior distribution. Thus, the
relaxation equations based on the maximization of the production of
generalized entropy (Sec. \ref{SectionAlgo}) can also provide a
parameterization of axisymmetric turbulence in the presence of a
small-scale forcing
\cite{Chavanis03,prior,kupka}.

\section{Dynamical stability of axisymmetric flows}
\label{SectionDyn}

\subsection{The axisymmetric Euler equations}
\label{SectionAxi}

The Euler equations describing the dynamics of an inviscid incompressible
axisymmetric flow can be written in cylindrical
coordinates $(r,\theta,z)$ as:
\EQA
\label{basic}
\frac{1}{r}\partial_r (ru)+\partial_z w&=&0 \; ,\\
\partial_t u+u\partial_r u+w\partial_z
u-\frac{v^2}{r}&=&-\frac{1}{\rho}\partial_r p \nonumber \; ,\\
\partial_t v+u\partial_r v+w\partial_z v+\frac{v u}{r}&=&0 \; ,\nonumber\\
\partial_t w+u\partial_r w+w\partial_z w&=&-\frac{1}{\rho}\partial_z p \; ,
\nonumber
\ENA
where $(u,v,w)$ denote the components of the velocity in a cylindrical
referential. Note that the third equation expresses the conservation
of the angular momentum $\sigma=r v$. The two other equations for $u$
and $w$ involve a pressure field determined through incompressibility. However, it can be eliminated by using the
stream-function vorticity formulation \cite{Lopez90}. The two new
scalar variables are the azimuthal component of the vorticity $\omega_{\theta}=\partial_{z}u-\partial_{r}w$ and the
stream function $\psi$ defined by:
\EQ
u = -\frac{1}{r}\partial_z \psi,\nonumber \qquad \text{and} \qquad w
= \frac{1}{r}\partial_r \psi \; .
\label{stream}
\EN
The existence of a stream function results from the incompressibility
and the axisymmetry of the flow. In this formulation, the system
(\ref{basic}) can be rewritten:
\EQA
\label{basic2}
\partial_t v -\frac{1}{r}\partial_z \psi \partial_r v + \frac{1}{r} \partial_r \psi  \partial_z v - \frac{1}{r^2} \partial_z \psi v &=&0 \; , \\ \nonumber
\partial_t \omega_\theta -\frac{1}{r}\partial_z \psi \partial_r \omega_\theta + \frac{1}{r}\partial_r \psi \partial_z
\omega_\theta + \partial_z \psi\frac{\omega_\theta}{r^2} &=&  \partial_z \left(
\frac{v^2}{r} \right) \;.
\ENA
By definition, the azimuthal component of the vorticity  is related to the stream function by:
\EQ
\partial_r\left(\frac{1}{r}\partial_r
\psi\right)+\frac{1}{r}\partial^2_z\psi=-\omega_\theta \; .
\label{beau}
\EN

We now introduce two new fields, the angular momentum
$\sigma=r v$ and $\xi$ which is related to the azimuthal component of the vorticity by $\xi=\omega_\theta/r$. Changing variables from $(r,z)$ to $(y,z)$ where $y=r^2/2$, we can finally recast the equations (\ref{basic2}) and (\ref{beau}) as
\EQA
\label{beautiful}
\partial_t \sigma+\lbrace \psi,\sigma\rbrace&=&0 \; , \\ \nonumber
\partial_t \xi+\lbrace
\psi,\xi\rbrace &=& \partial_z \left(\frac{\sigma^2}{4y^2}\right) \;
, \\ \nonumber
\Delta_{*}\psi\equiv \frac{1}{2y}\partial^2_z\psi+\partial^2_y\psi
&=& -\xi \; ,
\ENA
where
$\lbrace\psi,\phi\rbrace=\partial_{y}\psi\partial_{z}\phi-\partial_{z}
\psi\partial_{y}\phi$ is the Jacobian and $\Delta_{*}$ is a
pseudo-Laplacian. We also note that $u_z=\partial_y\psi=w$ and
$u_y=-\partial_z\psi=ru$. This formulation of the axisymmetric
Navier-Stokes equation has to be supplemented by appropriate
boundary conditions. For reasons which will become clear later, we delay this topic until the discussion of the conservation laws (section
\ref{SectionConservation} and Appendix \ref{AnnexeA}). In the
following, we study axisymmetric equilibrium flows by using the system
of equations (\ref{beautiful}) instead of (\ref{basic}). We look for
stationary solutions and investigate their stability by using
variational methods. Notice that only two scalar variables are
sufficient to prescribe such flows: we use $\sigma$, related to the
azimuthal component of the velocity field and $\xi$, related to the
azimuthal component of the vorticity.

\subsection{Stationary solutions}
\label{SectionStationary}

We  now derive the general form of stationary solutions of
the axisymmetric Euler equations (\ref{beautiful}).  Noting that
\EQA
\lbrace
\frac{\sigma}{2y},\sigma\rbrace=-\partial_z\left(\frac{\sigma^2}{4y^2}
\right) \; ,
\label{tr1}
\ENA
the stationary equations can be written
\EQA
\lbrace \psi,\sigma\rbrace = 0  \qquad \text{and} \qquad
\lbrace \psi,\xi\rbrace+\lbrace \frac{\sigma}{2y},\sigma\rbrace = 0.
\label{tr2}
\ENA
The first equation is satisfied if
\EQA
\psi=R(\sigma),
\label{tr3}
\ENA
where $R$ is an arbitrary function. Using the general identity
\EQA
\lbrace
R(\sigma),\xi\rbrace=R'(\sigma)\lbrace\sigma,\xi\rbrace=\lbrace\sigma,
\xi R'(\sigma)\rbrace,
\label{tr4}
\ENA
the second equation becomes
\EQA
\lbrace\sigma,\xi R'(\sigma)\rbrace+\lbrace \frac{\sigma}{2y},\sigma\rbrace=0,
\label{tr5}
\ENA
or, equivalently
\EQA
\lbrace\sigma,\xi R'(\sigma)-\frac{\sigma}{2y}\rbrace=0.
\label{tr6}
\ENA
Therefore, the general stationary solution of Eqs. (\ref{beautiful})
is of the form
\EQ
\psi = R(\sigma) \quad \text{and} \quad \xi
R'(\sigma)-\frac{\sigma}{2y}=G(\sigma),
\label{tr7}
\EN
where $R$ and $G$ are arbitrary functions. If $R$ is monotonic, we
can set $f=R^{-1}$ and we get $\sigma=f(\psi)$ and
\EQA
\xi- \frac{f(\psi)}{2y}\frac{1}{ R'\lbrack R^{-1}(\psi)\rbrack}=g(\psi).
\label{tr8}
\ENA
Using the identity
\EQA
\frac{1}{ R'\lbrack R^{-1}(\psi)\rbrack}=f'(\psi),
\label{tr9}
\ENA
we finally obtain
\EQA
\sigma=f(\psi),\nonumber\\
-\Delta_{*}\psi=\xi= \frac{f(\psi)}{2y}f'(\psi)+g(\psi),
\label{tr10}
\ENA
where $f$ and $g$ are arbitrary functions. We can obtain these
equations directly if we note that Eq. (\ref{tr2}-a) is satisfied if
$\sigma=f(\psi)$. Then, using  the general identity
\EQA
\biggl\lbrace \frac{\sigma}{2y},f(\psi)\biggr\rbrace=\biggl\lbrace
\frac{f'(\psi)\sigma}{2y},\psi\biggr\rbrace,
\label{tr11}
\ENA
we can rewrite Eq. (\ref{tr2}-b) in the form
\EQA
\biggl\lbrace \psi,\xi- \frac{f(\psi)f'(\psi)}{2y}\biggr\rbrace=0,
\label{tr12}
\ENA
which leads to Eq. (\ref{tr10}-b). Equation (\ref{tr10}-b) is the
fundamental differential equation of the problem which must be
supplemented by appropriate boundary conditions. Some particular
solutions of this equation will be given in
Sec. \ref{SectionSolution}. We will first show that the stationary
solutions can be found by a variational principle depending only on
the conservation laws of the system.

\subsection{Conservation laws}
\label{SectionConservation}

Axisymmetric inviscid flows satisfy a number of conservation laws. We
here give the expression of these conserved quantities and postpone
corresponding proofs in Appendix \ref{AnnexeA}. To derive the two
first conservation laws, we must assume that the function $\psi$
vanishes on the boundary of the domain which amounts to considering
that the normal component of the velocity is zero at the
boundary. This condition is not sufficient for deriving the third
conservation law and one must also suppose that either $\sigma$ or
$\xi$ vanishes at the boundary.

\begin{itemize}
\item The first conserved quantity is the total energy
\EQA
\label{ener}
E&=&\frac{1}{2}\int (u^2+w^2) \, rdrdz+\frac{1}{2}\int v^2 \, rdrdz
\\ \nonumber
&=&\frac{1}{2}\int \xi\psi \, dy dz + \frac{1}{4} \int
\frac{\sigma^2}{y} \, dydz.
\ENA
Here, we have normalized the energy by $2\pi$ and used integration by
parts to obtain the second expression.
\item Because of (\ref{beautiful}-a) any function of the angular
momentum is also an invariant  noted as
\begin{equation}
I_{f}=\int f(\sigma) \, dy dz \; .
\label{moments}
\end{equation}
These functionals are called the Casimirs. The conservation of all the
Casimirs is equivalent to the conservation of all the moments of
$\sigma$, denoted $I_{n}=\int \sigma^{n} \, dy dz$. The first moment
is the total angular momentum $I=\int \sigma \, dy dz$. \\ If
$\sigma=0$ or, more generally, $\sigma=\sigma(y)$, then $\xi$ is
conserved via (\ref{beautiful}-b). In that case, $\xi$ is called a
potential vorticity (or a pseudo-vorticity) and there is an additional
class of Casimir invariants: $I_{h}=\int h(\xi) \, dy dz$.  We
ignore this difficulty linked to a sort of ``degeneracy'' for the
time being.  Note that the situation where only the pseudo-vorticity is
conserved (i.e the case $\sigma=0$) has been treated in
\cite{Mohseni01}. In that case, the generalization essentially amounts
to replacing the Laplacian $\Delta$ in pure 2D flows by the
pseudo-Laplacian $\Delta_{*}$. The situation that we consider here is
complicated by the existence of additional invariants such as helicity
discussed below. This makes our situation intermediate between pure 2D
turbulence and 3D turbulence,  an interesting feature of our model.

\item Like in any 3D flows, the total helicity, $H = \int {\bf v}
\cdot \boldsymbol{\omega} \, rdrdz = \int \sigma \xi \, dy dz , \;$
is also an invariant. However, more
generally, we show in Appendix \ref{AnnexeA} the conservation of a
generalized helicity
\EQA
H_F=\int \xi F(\sigma) \, dy dz \; ,
\label{helicityn}
\ENA
where $F$ is an arbitrary function. In particular, the total
vorticity $H_0=\Gamma=\int \xi \, dy dz$ is conserved.\\
\end{itemize}

\subsection{Nonlinear dynamical stability}
\label{SectionDynstab}
From the integral constraints discussed previously, we can build a
functional ${\cal F}=E+I_{f}+H_{F}$. This functional is an invariant
of the inviscid dynamics. This is an extension of the Energy-Casimir
functional considered in \cite{Holm85}. It is also similar to a free
energy in thermodynamics. We now show that a critical point of ${\cal
F}$ determines a stationary solution of the axisymmetric Euler
equations. Furthermore, following \cite{Holm85}, a minimum or a
maximum of ${\cal F}$ provides a condition of formal nonlinear
dynamical stability. This means that a perturbation will remain close (in
some norm) to this minimum or maximum. Writing
\EQ
\delta {\cal F}=\delta (E+I_{f}+H_{F})=0,
\label{i1}
\EN
and taking variations on $\sigma$ and $\xi$, we obtain
\EQ
\psi+F(\sigma)=0
\quad \text{and} \quad
\frac{\sigma}{2y}+f'(\sigma)+\xi F'(\sigma)=0 \, .
\label{i3}
\EN
Setting $R=-F$ and $G=f'$, we recover the equations (\ref{tr7})
characterizing a steady solution of the axisymmetric Euler
equations. Since we obtain the general form of steady states it means
that we have found all the conservation laws of the axisymmetric Euler
equations.

In order to gain some physical insight in the problem, we
consider from now on a simpler model where only the usual helicity $H$
and the total vorticity $\Gamma$ are conserved instead of all the
generalized helicities. This is similar to our choice of restricting
ourselves to the Chandrasekhar model in axisymmetric MHD \cite{MHD}.
We define
\EQ
S[\sigma]=-\int C(\sigma) \, dy dz \, ,
\label{entropy2}
\EN
where $C$ is an arbitrary convex function, i.e. $C''>0$.  Such
functionals are exactly conserved by the axisymmetric equations (they
are particular Casimirs). Therefore, as in 2D hydrodynamics
\cite{Ellis02}, the maximization of $S$ at fixed energy
$E$, helicity $H$, circulation $\Gamma$ and angular momentum $I$
determines a nonlinearly dynamically stable stationary solution of the
axisymmetric Euler equations. This refined stability criterion is
stronger than the maximization of $J=S-\beta E-\mu H-\nu\Gamma-\alpha
I$ which just provides a {\it sufficient} condition of formal
nonlinear dynamical stability
\cite{Holm85}. The difference between these two criteria
is similar to a notion of ensemble inequivalence in thermodynamics
(where $S$ plays the role of an entropy and $J$ the role of a free
energy) \cite{Chavanis03,bb}. We shall not prove the nonlinear
dynamical stability result in this paper and refer to
\cite{Ellis02} for a precise discussion in 2D
hydrodynamics. In Sec. \ref{SectionSelective}, we show, however,
that this maximization principle is consistent with the phenomenology
of axisymmetric turbulence provided that $\sigma$ is interpreted as the {\it coarse-grained} angular momentum.

To first order, the variational
problem takes the form:
\EQ
\delta S-\beta\delta E-\mu\delta H-\gamma\delta\Gamma
-\alpha \delta I=0,
\label{variation}
\EN
where $\alpha$, $\beta$, $\mu$ and $\gamma$ are appropriate Lagrange
multipliers. This variational problem determining nonlinearly
dynamically stable stationary solutions of the Euler
equations is similar to a variational problem in thermodynamics where
$S$ plays the role of an entropy and $\beta$ the role of an inverse
temperature \cite{Chavanis03,prior}.  Using
the expression of $S$, $E$, $H$, $\Gamma$ and $I$, we find that the
solutions of (\ref{variation}) valid for any $\delta\sigma$ and
$\delta\xi$ satisfy
\EQA
\label{steady}
\beta\psi=-\mu\sigma-{\gamma}, \\ \nonumber
-C'(\sigma)={\beta}\frac{\sigma}{2y}+\mu\xi+\alpha,
\ENA
which is a particular case of Eq. (\ref{tr7}).  Thus, the variational
principle selects stationary solutions of the axisymmetric Euler
equations. We note that when only the ordinary helicity is considered
(instead of the general helicity), we obtain a linear relationship
between $\sigma$ and $\psi$. This is similar to the linear
relationship between velocity ${\bf V}$ and magnetic field ${\bf B}$
in MHD \cite{MHD}. Note that we have just considered the first order
variations here. To check if solutions (\ref{steady}) are real {\em
maxima} of $S$, one has to look for
second-order variations as discussed  in Appendix
\ref{VariationSecondes}.\

\subsection{$H$-functions and generalized selective decay principle}
\label{SectionSelective}

We now introduce the notion of fine-grained and coarse-grained
fields. The first one refers to the original field defined on all
points of space and time and the second one to a smooth version of it,
where a local average of the field has been performed. The coarse-grained field is also
defined on every point of space but contains less small-scale
structure than the original field. Since the functionals
(\ref{entropy2}) calculated with the fine-grained field $\sigma$ are 
particular Casimirs, they are rigorously conserved by the fine-grained
dynamics. In contrast, as Tremaine {\it et al.}
\cite{Tremaine86} have shown for the Vlasov equation in stellar
dynamics, the functionals of the form (\ref{entropy2}) calculated with
the coarse-grained field increase as mixing proceeds.  This is similar
to the Boltzmann $H$-theorem in kinetic theory except that the Vlasov
equation does not single out a particular functional. This property is
true also in the present context since Eq. (\ref{beautiful}-a) plays
the same role as the Vlasov equation. Therefore,
$S[\overline{\sigma}]=-\int C(\overline{\sigma})d{\bf r}$ increases
along the dynamics in the sense that $S[\overline{\sigma}({\bf
r},t)]\ge S[\overline{\sigma}({\bf r},0)]$ for all $C$ and all $t\ge
0$ where it is assumed that, initially, the flow is not mixed:
$\overline{\sigma}({\bf r},0)=\sigma({\bf r},0)$ (note that nothing is
implied concerning the relative values of $S(t)$ and $S(t')$ for $t\ge
0$, $t'\ge 0$). Following \cite{Tremaine86}, these functionals will be
called $H$-functions (or generalized $H$-functions). They also
increase (in that case monotonically) in the presence of viscosity
since the equations of motion now become:
\EQA
\label{EqVisqueuse}
\partial_t \sigma +\lbrace \psi, \sigma \rbrace&=& \nu r \biggl 
(\Delta \frac{\sigma}{r} - \frac{\sigma}{r^3}\biggr ) = \nu 
\biggl\lbrack \Delta \sigma - \frac{2}{r} \frac{\partial 
\sigma}{\partial r}\biggr\rbrack \; , \\ \nonumber
\partial_t \xi+\lbrace
\psi,\xi\rbrace &=& \partial_z \left(\frac{\sigma^2}{4y^2}\right) + 
\frac{\nu}{r} \biggl\lbrack\Delta (r \xi) - 
\frac{\xi}{r}\biggr\rbrack = \nu \biggl\lbrack \Delta \xi  + 
\frac{2}{r} \frac{\partial \xi}{\partial r}\biggr\rbrack \; ,
\ENA
and by integration by part, one can show that $\dot S=\nu \int 
C''(\sigma)(\nabla
\sigma)^2 d{\bf r}\ge 0$.  By contrast, the integrals $E$, $\Gamma$, $H$, and
$I$ are exactly or approximately conserved on the coarse-grained scale
(i.e. when they are calculated with the coarse-grained field) or in
the presence of a small viscosity. For example, in the presence of
viscosity, the kinetic energy evolves such that $\dot{E} = - \nu \int
\boldsymbol{\omega}^2 \, d{\bf r}$. It is easy to show that for
axisymmetric fields, the total vorticity $\boldsymbol{\omega} = r \xi
{\bf e_\theta} + {\bf \nabla} \times (\sigma / r \, {\bf e_\theta})$
vanishes in the long time limit. The demonstration is similar to
Cowling's \cite{Cowling34} theorem of dynamo theory which states that
an axisymmetric magnetic field cannot grow in an axisymmetric velocity
field: the first equation in (\ref{EqVisqueuse}) shows that $\sigma
\rightarrow 0$ for large time and, consequently, the source term in the
second equation $\partial_z \left(\sigma^2 / 4y^2\right)$ vanishes in
the long time limit, which implies that $\xi \rightarrow 0$. Thus, for
axisymmetric flows, both components of the vorticity vanish in the
long time limit and the energy is approximately conserved. In a
similar way, it can be shown that $\Gamma$, $H$, and $I$ are
approximately conserved and { must} therefore be strictly taken into
account in the constraints. Therefore, the functionals $S$ can be viewed
as {\it fragile invariants} while the constraints $E$, $\Gamma$, $H$,
and $I$ are {\it robust invariants}. This generalizes the notion of
selective decay in pure 2D turbulence where the enstrophy decays while
the energy is approximately conserved. In fact, minus the enstrophy is
a particular H-function
\cite{Chavanis03,prior,super}. The same discussion applies in the
present context. On the basis of this phenomenological principle, we
expect that, due to chaotic mixing and violent relaxation, the system
will reach a metaequilibrium state which maximizes a certain
$H$-function (non-universal) at fixed $E$, $\Gamma$, $H$, and
$I$. This phenomenological argument returns the variational principle
(\ref{variation}). Since this metaequilibrium state results from
turbulent mixing, it is expected to be particularly robust and should
possess properties of nonlinear dynamical stability. Therefore, the
stability arguments given previously are consistent with the
phenomenology of axisymmetric turbulence, provided that $\sigma$ is
interpreted as the coarse-grained angular momentum
$\overline{\sigma}$. This is remarkable because the two arguments are
relatively independent: there is no direct notion of decay (of $-S$)
in the first argument while this lies at the heart of the second.  In
fact, the phenomenology of violent relaxation explains {\it how} an
inviscid system can reach a nonlinearly dynamically stable stationary
state on a coarse-grained scale which is a maximum of a certain
$H$-function at fixed robust constraints (while $S[\sigma]$ is
rigorously conserved on the fine-grained scale). The point is that
during mixing $D\overline{\sigma}/Dt\neq 0$ and the $H$-functions
$S[\overline{\sigma}]$ increase. Once it has mixed
$D\overline{\sigma}/Dt= 0$ so that $\dot S[\overline{\sigma}]=0$. If
$\overline{\sigma}({\bf r},t)$ has been brought to a maximum
$\overline{\sigma}_{0}({\bf r})$ of a certain $H$-function (as a
result of mixing) and since $S[\overline{\sigma}]$ is conserved (after
mixing), then $\overline{\sigma}_{0}$ is a nonlinearly dynamically
stable stationary solution of the axisymmetric Euler equation
according to the stability criterion of Sec. \ref{SectionDynstab}.

\subsection{A numerical algorithm for the dynamical stability problem}
\label{SectionAlgo}

We shall construct a set of relaxation equations that increase
$S[\sigma]$ while conserving all the robust constraints $E$, $\Gamma$,
$H$, and $I$. These relaxation equations, which solve the optimization
problem of Sec. \ref{SectionDynstab}, can serve as powerful numerical
algorithm \footnote{The denomination ``numerical algorithm'' does not
exactly refer to its usual meaning. However, it has been used in
several articles on the subject \cite{Chavanis03,prior,kupka} to describe a
manner to solve an optimization problem (maximization of a functional
under constraints) by determining the stationary state of a relaxation
equation which can in turn be solved by classical numerical
methods. This
numerical algorithm has to be contrasted from the relaxation equations
of Secs. \ref{SectionMEPP} and \ref{SectionPrior} whose aim is to
provide a physical parameterization of turbulence out-of-equilibrium.} 
to compute arbitrary stationary solutions of the axisymmetric Euler
equations. In addition, they guarantee that these solutions are
nonlinearly dynamically stable with respect to the inviscid
dynamics. Such relaxation equations therefore have a clear practical
interest. They extend those obtained by Chavanis
\cite{Chavanis03,prior,kupka} in 2D hydrodynamics.

We write the dynamical equations as
\begin{eqnarray}
\label{relax2}
\frac{\partial \sigma}{\partial t}
=-\nabla\cdot {\bf J}_{\sigma}, \qquad
\frac{\partial\xi}{\partial t}
=
-\nabla\cdot {\bf J}_{\xi},
\end{eqnarray}
where ${\bf J}_{\sigma}$ and ${\bf J}_{\xi}$ are the currents to be
determined. We have not added advective terms since we here use these
equations just as numerical algorithms, not as a description of the
dynamics (see, however, Sec. \ref{SectionPrior}). By construction,
these equations satisfy the conservation of the total vorticity and
total angular momentum. On the other hand, the conservation of energy
and helicity impose the constraints
\EQA
\label{relax4}
\dot E&=&0=\int {\bf J}_{\xi}\cdot \nabla\psi \;  d{\bf r} \; + \;
\frac{1}{2}\int {\bf
J}_{\sigma}\cdot \nabla \biggl (\frac{\sigma}{y}\biggr ) \; d{\bf r}, \\
\dot H&=&0=\int {\bf J}_{\sigma}\cdot \nabla\xi \; d{\bf r} \; + \; \int {\bf
J}_{\xi}\cdot \nabla\sigma \; d{\bf r}.
\label{relax5}
\ENA
Finally, the time variations of $S[\sigma]$ are given by
\EQ
\dot S=-\int C''(\sigma) \; {\bf J}_{\sigma}\cdot \nabla\sigma \; d{\bf r}.
\label{relax3}
\EN
We  derive the optimal currents which maximize $\dot S$ with $\dot
E=\dot H=0$ and the additional constraints
\EQ
\frac{J_{\sigma}^{2}}{2} \le C_{\sigma}({\bf r},t) \qquad \text{and}
\qquad \frac{J_{\xi}^{2}}{2}\le C_{\xi}({\bf r},t),
\label{relax5b}
\EN
putting an upper bound on the currents. Writing the variational
principle in the form
\begin{eqnarray}
\delta \dot S-\beta(t)\delta \dot E-\mu(t)\delta \dot H-\int \chi \;
\delta \biggl (\frac{J_{\sigma}^{2}}{2}\biggr )d{\bf r}-\int \chi' \;
\delta \biggl (\frac{J_{\xi}^{2}}{2}\biggr )d{\bf r}=0,
\label{relax5c}
\end{eqnarray}
we obtain the optimal currents
\begin{eqnarray}
{\bf J}_{\sigma}&=&-D\biggl\lbrack \nabla\sigma + \frac{\beta}{C''(\sigma)}
\nabla\biggl (\frac{\sigma}{2y}\biggr )+ \frac{\mu}{C''(\sigma)}
\nabla\xi\biggr \rbrack,
\label{relax6}
\\ \nonumber
{\bf J}_{\xi}&=&-D'(\beta\nabla\psi+\mu\nabla\sigma),
\label{relax7}
\end{eqnarray}
where $\beta(t)$ and $\mu(t)$ are Lagrange multipliers which evolve in
time so as to conserve energy and helicity. They are determined by
substituting (\ref{relax6}) 
in the constraints
(\ref{relax4}) and (\ref{relax5}). Plugging the optimal
currents in (\ref{relax2}) we get
\begin{eqnarray}
\label{relax22}
\frac{\partial \sigma}{\partial t}
&=&\nabla\cdot \biggl\lbrace D\biggl\lbrack \nabla\sigma+\frac{\beta}
{C''(\sigma)}\nabla\biggl (\frac{\sigma}{2y}\biggr
)+\frac{\mu}{C''({\sigma})} \nabla\xi\biggr\rbrack\biggr \rbrace,
\\ \nonumber
\frac{\partial\xi}{\partial t}
&=&
\nabla\cdot  \biggl\lbrack
D'(\beta\nabla\psi+\mu\nabla\sigma)\biggr\rbrack.
\end{eqnarray}
It is straightforward to check that the system (\ref{relax22})
increases  the functional (\ref{entropy2}), {\em i.e.} $\dot S=\int
({\bf J}^{2}/D\rho +{\bf J}_{\xi}^{2}/D') d{\bf r}\ge 0$, and
that the stationary state is given by Eqs. (\ref{steady}).  Using the
same principle, we can also write relaxation equations which minimize
${\cal F}$. The optimal currents are
\begin{eqnarray}
{\bf J}_{\sigma}&=&-D\biggl\lbrack \nabla\sigma+\frac{1}{
f''(\sigma)}\nabla\biggl (\frac{\sigma}{2y}\biggr )+\frac{1}{
f''(\sigma)}\nabla(\xi F'(\sigma))\biggr \rbrack,
\label{relax6b}
\\ \nonumber
{\bf J}_{\xi}&=&-D'(\nabla\psi+\nabla F(\sigma)).
\label{relax7b}
\end{eqnarray}
and they return as an equilibrium state, the stationary solutions (\ref{i3}).

\subsection{Analytical solutions in simple cases}
\label{SectionSolution}
The steady state equations (\ref{tr10}) admit analytical solutions for simple shapes of the arbitrary functions $f$ and $g$. We will here derive some of these solutions and show that they are critical point of simple functionals.

\subsubsection{$g=0$}
Let us first consider the case where $g=0$. In that case, the steady
solution obeys
\EQ
\label{beautifulfin1}
\sigma = f(\psi) \quad \mathrm{and} \quad -\Delta_{*}\psi =
\xi=\frac{f(\psi)}{2y}f'(\psi) \; .
\EN
This equation admits simple solutions independent of $y$. Indeed, the
second equation becomes
\EQA
\frac{d^{2}\psi}{dz^{2}}=-\frac{1}{2}\frac{d}{d\psi}(f^{2}) \; ,
\label{aw1}
\ENA
which is equivalent to the motion of a particle in a potential
$\frac{1}{2}f^{2}(\psi)$ where $\psi$ plays the role of position and
$z$ the role of time. Multiplying both sides of Eq. (\ref{aw1}) by
$d\psi/dz$, and then integrating twice, the solution can be put under
parametric representation as
\EQ
z=\int^\psi \frac{d\phi}{\sqrt{K^2-f^2(\phi)}} \; ,
\label{solution1}
\EN
where $K$ is an integration constant and we have returned to original
variables. For example, for linear $f$, one obtains $\sigma \propto
\psi \propto \cos (K z)$.

\subsubsection{Constant g and linear $f$}

Consider now the case where $g$ is a constant $g = C$ and $f$ is a linear function of $\psi$, $f=A + B \psi$. The equations become
\EQA
\label{beautifulfin2} \sigma &=& A + B \psi \; , \\ \nonumber
-\Delta_{*}\psi &=& \xi = \frac{A B + B^2 \psi}{2y} + C \; .
\ENA
Note that these equations arise as critical points of the functional ${\cal F}_{0} = E + \mu_{0} H + \nu_{0} \Gamma + \alpha_{0} I$, i.e. they determine a state of minimum energy at fixed $H$, $\Gamma$ and $I$. Equation (\ref{beautifulfin2}-b) is an inhomogeneous linear
equation for $\psi$. The general solution is the sum of a special solution of the inhomogeneous equation superposed to the general solution of the homogeneous equation. A special solution is easily found as
\EQA
\label{SpecSol}
\psi &=& - \frac{A}{B} - 2 \frac{C y}{B^2} = - \frac{A}{B} - \frac{C
r^2}{B^2} \; , \\ \nonumber
\sigma &=& - 2 \frac{C y}{B}    \qquad \Rightarrow  \qquad v = -
\frac{C r}{B}  \; .
\ENA
This solution corresponds either to a laminar Taylor-Couette profile, or to a profile maximizing turbulent transport in Taylor-Couette flow \cite{Busse96}. Notice that the present theory is unable to  capture
the $1/r$ dependence of the Taylor-Couette flow because the solutions have to be regular at the origin. To reproduce such a behavior, one has to consider a domain with two boundaries (corresponding to the inner and outer cylinders), one  with $\psi=0$ and the other with $\psi \neq 0$. Such a procedure would introduce boundary terms in the
conserved quantities.

\begin{center}
\begin{figure}[h]
\includegraphics[scale=0.45,clip]{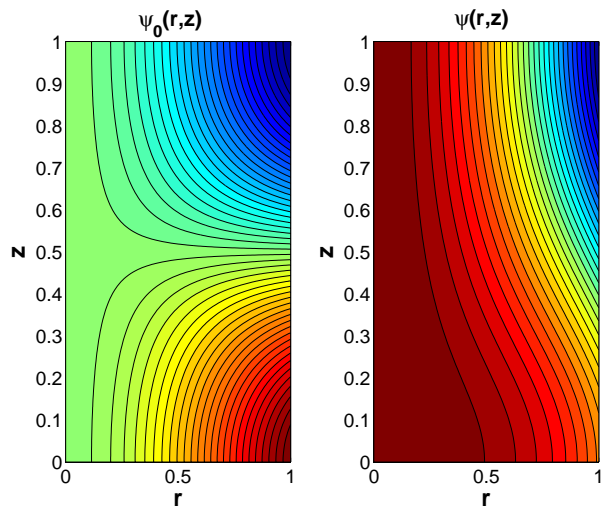}
\caption{\label{DessBeltrami} (Color online) Contour plot of the stream function associated to the \lq\lq Beltrami solution\rq\rq for the parameters $A=3$, $B=2$, $C=2$, $\mu=1/2$, $\kappa=\pi$, and $\phi=0$. The left hand side corresponds to the homogeneous solution (\ref{beltrami}) and the right hand side to the sum of the homogeneous solution and the particular solution (\ref{SpecSol}).}
\end{figure}
\end{center}

A general solution of the homogeneous equation can
be found by the method of separation of variables by writing $\psi =
G(y) F(z)$. It is then easy to show that $F(z) \propto \cos(\kappa z
+ \phi)$ where $\kappa$ and $\phi$ are two constants. Then, one finds
that $G$ obeys the following equation:
\EQ
\frac{d^2 G}{dy^2}+\left(\frac{B^2-\kappa^2}{2y}\right) G =0 \quad
\Leftrightarrow \quad
r^2 \frac{d^2}{dr^2}\left[\frac{G}{r}\right] + r
\frac{d}{dr}\left[\frac{G}{r}\right] +
\left[(B^2-\kappa^2)r^2-1\right]\frac{G}{r} = 0 \; ,
\label{Bessel}
\EN
whose solution can be expressed in term of Bessel function of first order. The general solution for $\psi$ is thus
\EQ
\psi_0 = \mu r J_1\left(\sqrt{(B^2-\kappa^2)} r \right) \cos(\kappa z+\phi) \; ,
\label{beltrami}
\EN
where $\mu$, $\kappa$ and $\phi$ are integration constants. This
solution is a critical point of the functional ${\cal
F}_{B}=E+\mu_{0}H$, i.e. it minimizes the energy at fixed Helicity. In
vectorial form, this leads to ${\bf curl}({\bf u})=\lambda {\bf u}$
such that vorticity and velocity are everywhere proportional. This is
the so-called Beltrami solution which has proven to be important in
the study of the dynamo effect \cite{Kageyama99}, \ie the generation
of a magnetic field by a conducting fluid. The most popular flow of
this type is the Roberts flow \cite{Roberts72}. In the limit $\kappa
\to B$, the homogeneous solution tends to $r^2$ and this solution becomes
equivalent to one of the nonlinear self-similar solution of the von
K\'arm\'an flow found by Zandbergen and Dijkstra
\cite{Zandbergen87}. On figure \ref{DessBeltrami}, we show a contour
plot of this solution.

\subsubsection{$f$ and $g$ linear}

\begin{center}
\begin{figure}[h]
\includegraphics[scale=0.35,clip]{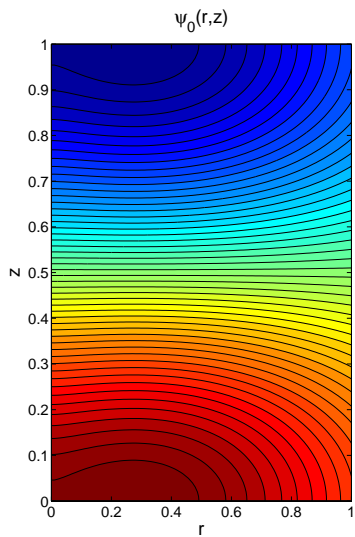}
\caption{\label{DessWhitt} (Color online) Contour plot of the stream function associated to the \lq\lq Whittaker solution\rq\rq of equation (\ref{Whittaker}) for the parameters $A=3$, $B=2$, $C=2$, $D=-3$, $\mu=1/2$, $\kappa=\pi$ and $\phi=0$.}
\end{figure}
\end{center}

The case where both $f$ and $g$ are linear $g=C+D\psi$ and
$f=A+B\psi$ is similar to the previous one. The equations are now
\EQA
\sigma &=& A+B\psi ,\nonumber\\
-\Delta_{*}\psi&=&\xi= \left[\frac{B^2}{2y} + D \right] \psi +
\frac{A B}{2y}+C \; .
\label{beautifulfin21}
\ENA
Note that these equations are obtained by minimizing the second
moment of angular momentum $I_{2}=\int \sigma^{2} dy dz$ at fixed $E$,
$H$, $\Gamma$ and $I$. They represent, therefore, the counterpart of
the minimum enstrophy principle in 2D hydrodynamics, leading to a
linear relationship between vorticity and stream function. Equation
(\ref{beautifulfin21}-b) is an inhomogeneous linear equation for
$\psi$. The general solution is the sum of a special solution of the
inhomogeneous equation superposed to the general solution of the
homogeneous equation. Solutions of the homogeneous equation can be
found by assuming separation of variable as previously $\psi=F(z)G(y)$. The
solution for $F$ is $F(z)=\cos(\kappa z+\phi)$, where $\kappa$
and $\phi$ are two integration constants. Then, the equation for $G$
involves a supplementary term compared to the one in the previous
section:
\EQ
\frac{d^2 G}{dy^2}+\left(\frac{B^2-\kappa^2}{2y}+D\right) G =0.
\label{Whittaker}
\EN
The two solutions of this equations can be expressed in terms of
Whittaker function (see \cite{Gradshteyn65}, p.~1059) $W_{\lambda,\pm
1/2}(2\sqrt{-D}y)$,
$\lambda=(B^2-\kappa^2)/8D$. These function behave at infinity like
$y^\lambda \exp(-y/2)$. Turning back to original variable, one can
therefore express the general solution of equation
(\ref{beautifulfin21}) as
\EQ
\psi_0 = \mu W_{\lambda,\pm 1/2}(\sqrt{-D}r^2)\cos(\kappa z+\phi),
\label{Whittaker2}
\EN
where $C$ is an integration constant. Note that the negativeness of the coefficient $D$ is imposed by an asymptotical analysis of equation (\ref{Whittaker}): when $y \rightarrow \infty$, we see that a positive coefficient $D$ would introduce an oscillatory, unphysical, behavior for $G$. Figure \ref{DessWhitt} shows a typical realization of this solution.

\section{Statistical mechanics of axisymmetric flows}
\label{SectionStat}

\subsection{The mixing entropy}
\label{SectionEntropy}

Starting with a given initial condition, equations (\ref{beautiful}) are expected to develop a complicated mixing process, with formation of finer and finer structure, leading to more and more degrees of freedom. A precise prediction of the state of the system would a priori require to keep track of all these degrees of freedom. Suppose however that we are only interested in the knowledge of the system at some coarse-grained scale. Since mixing is continuously occurring at smaller and smaller scales, we can expect the formation of a {\it
metaequilibrium} state on the coarse-grained scale. Our goal is to derive its shape through thermodynamical arguments, based on entropy maximization in the spirit of \cite{Jaynes57}. We focus here on basics. More discussion about this procedure can be found in \cite{Robert91}. We introduce coarse-grained and fine-grained quantities. According to Eq. (\ref{beautiful}-c), $\psi=\Delta_*^{-1}\xi$ is expressed as an integral
over $\xi$. It is therefore a smooth function whose fluctuations can be neglected $\psi = \overline{\psi}$. We shall determine the distribution of fluctuations
of angular momentum $\sigma$ by an approach similar to that developed in 2D  turbulence. We then introduce $\rho({\bf
r},\eta)$, the density probability of finding the value $\sigma=\eta$ at position ${\bf r}$. Then, the coarse-grained angular momentum is $\overline{\sigma}=\int\rho\eta d\eta$ and the local normalization $\int \rho d\eta =1$.  We introduce the mixing entropy
\begin{equation}
S[\rho]=-\int \rho\ln \rho \; dy\, dz \;  d\eta \; .
\label{entropym}
\end{equation}
This functional $S[\rho]$ is equal to the logarithm of the disorder
where the disorder is the number of microstates corresponding to the
macrostate $\rho({\bf r},\eta)$ which can be obtained by a combinatorial analysis (see \cite{super} for details). The fact that the entropy depends
only on the distribution of angular momentum guarantees that it will
be conserved by the fine-grained dynamics (however, it increases on a coarse-grained scale as we will show subsequently). As a drawback, we will not
be able to characterize the fluctuations of pseudo-vorticity $\xi$ by
this method; we will simply get its coarse-grained value. This
difficulty may reflect the fact that we are in a situation
intermediate between 2D and 3D turbulence. In 2D turbulence, the
distribution of vorticity is enough to construct equilibrium solutions
whereas, in 3D turbulence, it is well-known that the average of
fluctuating quantities (such as the Reynolds stress) are very
important. In Appendix \ref{fluc}, we consider another approach which
puts the fluctuations of $\sigma$ and $\xi$ on an equal footing. This
approach seems to indicate that the fluctuations of $\xi$ have a
peculiar behavior that may give rise to a sort of \lq\lq phase
transition\rq\rq. Here, for simplicity and clarity, we first
concentrate on the simplified situation (similar to 2D turbulence)
where the fluctuations of $\xi$ are mild; however, we keep in mind
that there may be another regime (closer to 3D turbulence)
characterizing axisymmetric flows.

The coarse-grained values of the constraints can be written as
\EQA
\overline{E} &=& \frac{1}{2} \int \xi \psi \, dy dz \;  +  \;
\frac{1}{4} \int \frac{\overline{\sigma}^{2}}{y} \, dy dz \; ,
\label{energy2}
\\
\overline{H} &=& \int \xi \, \overline{\sigma} \, dy dz  \; , \qquad
\Gamma = \int \xi \, dy dz \; ,
\label{helicity2}
\\
&&\overline{I_{n}} = \int \rho \eta^{n} \; dy dz \,  d\eta \; .
\label{moments2}
\ENA
For the coarse-grained energy, we made the non trivial hypothesis that
the fluctuations of the energy could be neglected and used the
coarse-grained field $\overline{\sigma}$ to calculate the mean
energy. This is justified by the remark made in section
\ref{SectionSelective} that in the presence of a small viscosity (or
 a coarse-graining), the energy is approximately
conserved.

As indicated previously, $E$, $\Gamma$, $H$ and $I$ are
robust constraints. Thus they can be determined at any time since
their exact value is close to their coarse-grained value. They have
thus been expressed directly in terms of the coarse-grained field. By
contrast the higher moments $I_{n>1}$ are fragile constraints because
they are altered by coarse-graining.  They can be determined only from
the initial conditions which are supposed un-mixed (or from the
fine-grained field) since they are affected by coarse-graining in the
sense that $I_{n>1}[\overline{\sigma}]=\int
\overline{\sigma}^{n} dy dz \neq \overline{I_{n>1}[{\sigma}]}=\int
\overline{\sigma^{n}} dy dz =\int \rho
\eta^{n} dy dz d\eta$. Part of the Casimirs goes into the
coarse-grained field and part goes into fine-grained fluctuations.
In a sense, $I_{n>1}$ are ``hidden constraints'' because they
cannot be determined from the coarse-grained field. Therefore, the
statistical theory assumes that we know the initial conditions in
detail and that these initial conditions represent the fine-grained
field. This poses a conceptual problem  because in real situations
we are not sure whether the initial condition is already mixed or
not.  Furthermore, if the initial condition already results from a
mixing process (like vortices formed in a succession of mergings in 2D
decaying turbulence), it is more proper to ignore its fine structure
and take its coarse-grained density as a new ``fine-grained'' initial
condition to predict the next merging (see \cite{JFM1}, p 284). In fact, due
to viscosity, the fine structure of the field is progressively
erased. These difficulties are intrinsic to the statistical theory of
continuous fields.

\subsection{The Gibbs state}
\label{SectionGibbs}

The most probable distribution at metaequilibrium is obtained by
maximizing the mixing entropy $S[\rho]$ at fixed $\overline{E}$,
$\overline{H}$, $\overline{\Gamma}$, $\overline{I}$,
$\overline{I_{n}}$ and normalization. We introduce Lagrange
multipliers and write the variational principle in the form
\begin{eqnarray}
\label{vp}
\delta S - \beta\delta \overline{E} - \mu\delta \overline{H} - \gamma
\delta \overline{\Gamma} - \alpha \delta \overline{I}
- \sum_{n>1}\alpha_{n}\delta \biggl ( \int \rho \eta^{n} \, dy dz 
d\eta\biggr ) \\ \nonumber-\int \zeta(y,z) \delta \biggl (\int \rho
d\eta
\biggr
)\, dy dz = 0 \; .
\end{eqnarray}
The last term in this equation corresponds to the normalization of the probability density in each point of space and thus needs the introduction of one Lagrange multiplier $\zeta(y,z)$ for each point $(y,z)$. We shall treat the variations on $\rho$ and $\xi$ independently. The
variations on $\xi$ imply
\begin{equation}
\beta\psi = -\mu  \overline{\sigma } - \gamma.
\label{gibbsm}
\end{equation}
The variations on $\rho$ yield the Gibbs state
\begin{equation}
\rho = \frac{1}{Z(y,z)} \chi(\eta)e^{-(\beta\frac{\overline{\sigma}}{
2y}+\mu\xi+\alpha)\eta} \; ,
\label{rhoz}
\end{equation}
where
\begin{equation}
\chi(\eta)=e^{-\sum_{n>1} \alpha_{n}\eta^{n}}.
\label{chii}
\end{equation}
To prepare the approach of Sec. \ref{SectionPrior}, we have
distinguished between the Lagrange multipliers $\alpha_{n>1}$ which
account for the conservation of the fragile constraints $I_{n>1}$
(they have been regrouped in the function $\chi(\eta)$) from the
Lagrange multipliers $\beta$, $\mu$ and $\alpha$ which are related to
the robust constraints. Therefore, the distribution (\ref{rhoz}) is
the product of a universal Boltzmann factor and of a non-universal
function $\chi(\eta)$ depending on the initial conditions. Note that
instead of conserving the fine-grained moments $I_{n}=\int
\overline{\sigma^{n}}dydz=\int
\rho\eta^{n}d\eta dy dz$ of angular momentum, we can equivalently
conserve the total area $\gamma(\eta)=\int\rho dydz$ of each level.
The ``partition function'' is determined by the local normalization
condition yielding
\begin{equation}
Z=\int \chi(\eta)e^{-(\beta\frac{\overline{\sigma}}{2y}+\mu\xi+\alpha)\eta} d\eta.
\label{partition}
\end{equation}
We note that $Z$ is a
function of
\begin{equation}
\Psi=\beta\frac{\overline{\sigma}}{2y}+\mu\xi+\alpha,
\label{partition2}
\end{equation}
expressed by a sort of generalized Laplace transform (this is not
exactly a Laplace transform since the variable $\eta$ can take
positive and negative values):
\begin{equation}
Z(\Psi)=\int_{-\infty}^{+\infty} \chi(\eta)e^{-\eta\Psi} d\eta \equiv
\hat{\chi}(\Psi).
\label{partition3}
\end{equation}
The coarse-grained angular momentum is given by
\begin{equation}
\overline{\sigma}=\frac{1}{ Z(\Psi)}\int \eta\chi(\eta)e^{-\eta\Psi} d\eta.
\label{partition4}
\end{equation}
It is straightforward to establish that
\begin{equation}
\overline{\sigma}=-\frac{\partial\ln Z}{\partial \Psi}=F(\Psi) \qquad
\text{and} \qquad
\sigma_{2}=\frac{\partial^{2}\ln Z}{\partial\Psi^{2}}=-F'(\Psi),
\label{var}
\end{equation}
where
\begin{equation}
\sigma_{2}=\overline{(\sigma-\overline{\sigma})^{2}}=\overline{\sigma
^{2}}-\overline{\sigma}^{2} \; ,
\label{varia}
\end{equation}
is the centered variance of the local
angular momentum distribution (we have noted $\overline{\sigma^{n}}=\int
\rho\eta^{n}d\eta$). According to Eq. (\ref{var}-b), $F$ is
monotonically decreasing. Therefore, relation (\ref{var}-a) can be
inverted and we get
\EQA
\beta\psi&=&-\mu\overline{\sigma}-\gamma \; , \label{gibbsre}
\\
\beta\frac{\overline{\sigma}}{ 2y}&+&\mu\xi+\alpha=F^{-1}(\overline{\sigma}) \; .
\label{gibbsty}
\ENA
Comparing with Eq. (\ref{tr7}), we check explicitly that the
coarse-grained flow is a stationary solution of the axisymmetric Euler
equations. Therefore, for given initial conditions, the statistical
theory selects a particular stationary solution among all possible
ones. We remark that the differential equation for $\psi$, arising
from Eq. (\ref{gibbsty}) and (\ref{beautiful})-c involves the {\it
inverse} $F^{-1}$ of the function determined by Eq. (\ref{var}-a)
while in pure 2D turbulence it involves the direct function $F$,
i.e. $-\Delta\psi=F(\beta\psi+\alpha)$. This ``inversion'' is another
striking particularity of our system.

Comparing Eq. (\ref{gibbsty}) with Eq. (\ref{steady}), we note that
the equilibrium coarse-grained angular velocity maximizes a certain
H-function where $C$ is related to $F$ by $F^{-1}(\overline{\sigma})=-C'(\overline{\sigma})$, i.e.
\begin{equation}
C(\overline{\sigma})=-\int^{\overline{\sigma}} F^{-1}(x)dx \; .
\label{cf}
\end{equation}
The H-function $S[\overline{\sigma}]=-\int C(\overline{\sigma})d{\bf
r}$ selected by the statistical theory can be viewed as a
``generalized entropy'' in the reduced $\overline{\sigma}$-space
\cite{Chavanis03,prior,kupka,super}. It depends on the initial conditions
through the function $\chi(\eta)$ which must be related to the
fine-grained moments of angular momentum (Casimirs). Therefore, in
this approach where the constraints associated with the Casimirs are
treated microcanonically, the generalized entropy in $\overline{\sigma}$-space
can only be obtained {\it a posteriori}, after having solved the full
equilibrium equations and related the Lagrange multipliers to the
constraints. 

Using $\sigma_{2}=- \, d\,
\overline{\sigma}/d\Psi$ according to (\ref{var})-b and $\Psi=-C'(\overline{\sigma})$ according to (\ref{steady})-b, we get the
identity
\begin{equation}
\sigma_{2}=\frac{1}{ C''(\overline{\sigma})}.
\label{sfw}
\end{equation}
Therefore, at equilibrium, there is a functional relation between the
variance $\sigma_{2}$ of the distribution and the coarse-grained 
angular momentum $\overline{\sigma}$ through the second derivative of
the function $C$. This is similar to the ``fluctuation-dissipation''
theorem \cite{sw}. Finally, we note that the most probable value $\langle \sigma\rangle(y,z)$
of the distribution $\rho(y,z,\eta)$ is such that
${\cal F}(\eta)= -\Psi\eta+\ln\chi(\eta)$ is maximum yielding $(\ln\chi)''<0$ and
\begin{equation}
\langle \sigma\rangle=\lbrack (\ln\chi)'\rbrack^{-1}(\Psi)=G(\Psi),
\label{mp1}
\end{equation}
where $G$ is monotonically decreasing. In general, the most probable value $\langle\sigma\rangle$ of the distribution (\ref{rhoz}) does not coincide with the mean value $\bar\sigma$. The condition $\langle\sigma\rangle=\bar\sigma$ is equivalent to
\begin{equation}
-(\ln\hat\chi)'=\lbrack (\ln\chi)'\rbrack^{-1}.
\label{mp2}
\end{equation}
This equality holds if $\chi$ is Gaussian.  Furthermore, we show in
Appendix \ref{sec_Z} that $\langle
\sigma\rangle$ is a stationary solution of the axisymmetric Euler
equations only when $\langle\sigma\rangle=\overline{\sigma}$.

\subsection{Particular cases}
Some particular cases of $F(\Psi)$ relationships have been collected
in \cite{Chavanis03,prior,super}.  We shall specify different forms of
function $\chi(\eta)$ and determine the corresponding $F(\Psi)$ and
$S\lbrack\overline{\sigma}\rbrack$ from the preceding relations. We
refer to \cite{Chavanis03} for more details. In the two-levels case
where $\eta=\lambda_{0},\lambda_{1}$, we get
\begin{equation}
\overline{\sigma} = F(\Psi) =
\lambda_{0}+\frac{\lambda_{1}-\lambda_{0}}{
1+e^{(\lambda_{1}-\lambda_{0})\Psi}}.
\label{ex1}
\end{equation}
In the present case, we need to invert this relation to express $\xi$
as a function of $\overline{\sigma}$, hence $\psi$. As discussed
above, this situation is reversed with respect to pure 2D plane
flows. We thus obtain
\begin{equation}
-\Delta_{*}\psi=\xi=\frac{1}{ \mu(\lambda_{1}-\lambda_{0})}\ln\biggl
(\frac{\mu\lambda_{1}+\beta\psi+\gamma}{-\beta\psi-\gamma-\lambda_{0}\mu}\biggr
)+\frac{1}{2}\frac{\beta^{2}}{\mu^{2}}\frac{\psi}{y}+\frac{\beta\gamma
}{2\mu^{2}y}-\frac{\gamma}{\mu}.
\label{ex2}
\end{equation}
In that case, the generalized entropy in $\overline{\sigma}$-space has the form
\begin{equation}
S[\overline{\sigma}]=-\int \lbrack p\ln p+(1-p)\ln (1-p)\rbrack d{\bf r} \; ,
\label{ex3}
\end{equation}
with $\overline{\sigma}=p\lambda_{1}+(1-p)\lambda_{0}$. This is
similar to the Fermi-Dirac entropy. In this two-levels case, the
generalized entropy $S[\overline{\sigma}]$ defined by
Eqs. (\ref{entropy2})-(\ref{cf}) coincides with the mixing entropy
$S[\rho]$ defined by Eq. (\ref{entropym}). This is the only situation
where we have this equivalence. Taking $\lambda_{0}=0$ and considering
the dilute limit $\overline{\sigma}\ll
\lambda_{1}$, we get
\begin{equation}
\overline{\sigma}=\lambda_{1}e^{-\lambda_{1}\Psi},
\label{ex1a}
\end{equation}
leading to
\begin{equation}
-\Delta_{*}\psi=\xi=-\frac{1}{ \lambda_{1}\mu}\ln\biggl
(\frac{-\beta\psi-\gamma}{\mu\lambda_{1}}\biggr
)+\frac{1}{2}\frac{\beta^{2}}{\mu^{2}}\frac{\psi}{y}+\frac{\beta\gamma
}{2\mu^{2}y}-\frac{\alpha}{\mu}  .
\label{ex2b}
\end{equation}
In that case, the generalized entropy in $\overline{\sigma}$-space is
similar to the  Boltzmann entropy
\begin{equation}
S[\overline{\sigma}]=-\int
\frac{\overline{\sigma}}{\lambda_{1}}\ln\frac{\overline{\sigma}}{\lambda_{
1}} d{\bf r}.
\end{equation}

If $\chi(\eta)$ is a Gaussian, then
\begin{equation}
\overline{\sigma}=-\sigma_{2}\Psi,
\label{ex4}
\end{equation}
where the centered variance $\sigma_{2}$ is a constant. Inverting
this relation, we get
\begin{equation}
-\Delta_{*}\psi=\xi=\frac{\beta}{\mu^{2}\sigma_{2}}\psi+\frac{\gamma}{
\mu^{2}\sigma_{2}}+\frac{1}{2}\frac{\beta^{2}}{\mu^{2}}\frac{\psi}{y}+
\frac{\beta\gamma}{2\mu^{2}y}-\frac{\alpha}{\mu}.
\label{ex5}
\end{equation}
This is the type of mean-field equations that we have considered in
Sec. \ref{SectionSolution}.  In that case, the generalized entropy in $\overline{\sigma}$-space is
\begin{equation}
S[\overline{\sigma}]=-\frac{1}{ 2\sigma_{2}}\int \overline{\sigma}^{2}d{\bf r},
\label{ex6}
\end{equation}
which is similar to minus the enstrophy in pure 2D hydrodynamics.

If $\chi(\eta)$ is a decentered Gamma distribution 
\cite{Ellis02}-\cite{super}, then
\begin{equation}
\overline{\sigma}=-\frac{\sigma_{2}\Psi}{1+\lambda\sigma_{2}\Psi} \; ,
\label{ex7}
\end{equation}
where the centered variance $\sigma_{2}$ is a constant and 
$2\lambda\sigma_{2}^{1/2}$ is equal to the skewness of $\chi(\eta)$. 
Inverting this relation, we get
\begin{equation}
-\Delta_{*}\psi=\xi=\frac{\beta\psi+\gamma}{\sigma_{2} \mu 
(1-\lambda\beta 
\psi-\lambda\gamma)}+\frac{1}{2}\frac{\beta^{2}}{\mu^{2}}\frac{\psi}{y}
+\frac{\beta\gamma}{2\mu^{2}y}-\frac{\alpha}{\mu} \; .
\label{ex8}
\end{equation}
In that case, the generalized entropy in $\overline{\sigma}$-space is
\begin{equation}
S[\overline{\sigma}]=-\frac{1}{ \lambda\sigma_{2}}\int \biggl\lbrack
\overline{\sigma}-\frac{1}{\lambda}\ln
(1+\lambda\overline{\sigma})\biggr\rbrack d{\bf r}.
\label{ex9}
\end{equation}

\subsection{Relaxation towards the statistical equilibrium state}
\label{SectionMEPP}

We would like now to construct a system of relaxation equations which
conserve all the invariants of the inviscid dynamics (robust and
fragile) and relax towards the statistical equilibrium state. These
equations can be used as a numerical algorithm to construct the
statistical equilibrium state. They also provide a subgrid scale
parameterization of axisymmetric turbulence. In that context, they can
describe the dynamical evolution of the flow on the coarse-grained
scale. Note that in the coarse-grained formulation, the inviscid 
approximation is easier to justify, since viscosity only acts at very
small scales. Following the approach of Robert \& Sommeria
\cite{Robert92}, these relaxation equations can be obtained by using a
Maximum Entropy Production Principle (MEPP).

The equations of evolution for the coarse-grained fields are given by
\begin{eqnarray}
\label{relax1}
\frac{\partial \overline{\sigma}}{\partial t} + \overline{\bf u}\cdot
\nabla \overline{\sigma}
&=& -\nabla\cdot {\bf J}_{\sigma} \; , \\ \nonumber
\frac{\partial\xi}{\partial t} + \overline{\bf u}\cdot \nabla {\xi} &=&
\frac{\partial}{\partial x}\biggl (\frac{{\overline{\sigma}}^{2}}{4 y^{2}}\biggr )
-\nabla \cdot {\bf J}_{\xi} \; ,
\end{eqnarray}
where ${\bf J}_\sigma$ and ${\bf J}_\xi$ are currents which contain all the
information coming from interaction with sub-grid scales. Note that we
have kept the advective terms because these equations are expected to
describe the relaxation of the flow (on the coarse-grained scale)
towards statistical equilibrium; they are not only numerical
algorithms. If we want to keep track of the conservation of all the
Casimirs (or equivalently of the total area of each level of angular
momentum), we need to introduce equations of conservation for the
density probability $\rho({\bf
r},\eta,t)$ of angular momentum. We write them as
\begin{eqnarray}
\frac{\partial \rho}{\partial t} + \overline{\bf u}\cdot \nabla \rho
=-\nabla\cdot {\bf J} \; , \label{relax20}
\end{eqnarray}
where ${\bf J}({\bf r},\eta,t)$ is the current of the level $\eta$ of angular
momentum. Multiplying Eq. (\ref{relax20}) by $\eta$ and integrating
over all the levels, we recover Eq. (\ref{relax1}-a) with ${\bf
J}_{\sigma}=\int {\bf J} \, \eta \; d\eta$. Furthermore, the conservation of
the local normalization $\int \rho \, d\eta=1$ imposes
\begin{eqnarray}
\int {\bf J} \; d\eta=0.
\label{relax51}
\end{eqnarray}
The time variations of $S[\rho]$ are given by
\begin{eqnarray}
\dot S=-\int {\bf J} \cdot \nabla\ln\rho \; d{\bf r}d\eta,
\label{relax3b}
\end{eqnarray}
while the time variations of $E$ and $H$ have been given previously in
Eqs. (\ref{relax4})-(\ref{relax5}).  Following the MEPP, we maximize
$\dot S$ with $\dot E=\dot H=0$, the normalization constraint
(\ref{relax51}) and the additional constraints
\begin{eqnarray}
\int \frac{J^{2}}{2\rho}d\eta\le C({\bf r},t), \qquad
\frac{J_{\xi}^{2}}{2}\le C_{\xi}({\bf r},t).
\label{relax52}
\end{eqnarray}
Writing the variational principle in the form
\begin{eqnarray}
\delta\dot S-\beta\delta\dot E-\mu\delta\dot H-\int\zeta\delta\biggl
(\int {\bf J}d\eta\biggr )d{\bf r}-\int {\chi}\delta\biggl
(\frac{J^{2}}{2\rho}\biggr )d{\bf r}d\eta-\int
{\chi'}\delta\biggl (\frac{J_{\xi}^{2}}{2\xi}\biggr )d{\bf
r}=0,
\label{relax53}
\end{eqnarray}
we obtain the optimal currents
\begin{eqnarray}
{\bf J} &=& -D\biggl\lbrack \nabla\rho+\frac{1}{
2}\beta\rho(\eta-\overline{\sigma})\nabla\biggl
(\frac{\overline{\sigma}}{y}\biggr
)+\mu\rho(\eta-\overline{\sigma})\nabla\xi\biggr \rbrack,
\label{relax54}
\\
{\bf J}_{\xi} &=& -D'(\beta\nabla\psi+\mu\nabla\sigma),
\label{relax55}
\end{eqnarray}
where we have used Eq. (\ref{relax51}) to obtain the final expression
of the current (\ref{relax54}). The current of angular momentum is
therefore given by
\begin{eqnarray}
{\bf J}_{\sigma} = -D\biggl\lbrack \nabla\overline{\sigma}+\frac{1}{
2}\beta\sigma_{2}\nabla\biggl (\frac{\overline{\sigma}}{y}\biggr
)+\mu\sigma_{2}\nabla\xi\biggr \rbrack,
\label{ca}
\end{eqnarray}
where $\sigma_{2}$ is defined in Eq. (\ref{varia}).  We note that
$\beta(t)$ and $\mu(t)$ are time dependent Lagrange multipliers that
evolve in order to conserve energy and helicity. Their explicit
expression is obtained by inserting Eqs. (\ref{relax55}) and (\ref{ca})
in the constraints (\ref{relax4})-(\ref{relax5}).

We now show that the relaxation equations (\ref{relax1}-b) and
(\ref{relax20}) with the currents (\ref{relax55}) and (\ref{relax54})
increase the mixing entropy (\ref{entropym}) until the Gibbs state is
reached ($H$-theorem). We can write the rate of entropy production
(\ref{relax3b}) in the form
\begin{eqnarray}
\label{relax60}
\dot S=-\int \frac{{\bf J}}{\rho}\cdot \biggl\lbrack \nabla\rho+\frac{1}{
2}\beta\rho(\eta-\overline{\sigma})\nabla\biggl
(\frac{\overline{\sigma}}{y}\biggr
)+\mu\rho(\eta-\overline{\sigma})\nabla\xi\biggr \rbrack \, d{\bf r}d\eta
\\ \nonumber
+\frac{1}{2}\beta \int (\eta-\overline{\sigma}) {\bf J}\cdot
\nabla\biggl (\frac{\overline{\sigma}}{y}\biggr ) \, d{\bf r}d\eta +\mu \int  (\eta-\overline{\sigma}){\bf
J}\cdot \nabla\xi
\, d{\bf r} d\eta.
\end{eqnarray}
Using Eqs. (\ref{relax55}) and (\ref{relax54})  this can be rewritten
\begin{eqnarray}
\dot S=\int \frac{{\bf J}^{2}}{D\rho} \, d{\bf r}
+ \frac{1}{2}\beta \int {\bf J}_{\sigma}\cdot \nabla\biggl
(\frac{\overline{\sigma}}{y}\biggr ) \, d{\bf r} +\mu \int {\bf J}_{\sigma}\cdot \nabla\xi \, d{\bf r}.
\label{relax61}
\end{eqnarray}
Using Eqs. (\ref{relax4})-(\ref{relax5}), we get
\begin{eqnarray}
\dot S=\int \frac{{\bf J}^{2}}{D\rho} d{\bf r}
- \int {\bf J}_{\xi}\cdot (\beta\nabla\psi+\mu\nabla\overline{\sigma})d{\bf r},
\label{relax62}
\end{eqnarray}
hence
\begin{eqnarray}
\dot S=\int \frac{{\bf J}^{2}}{D\rho} d{\bf r}+\int \frac{{\bf
J}_{\xi}^{2}}{D'} d{\bf r}.
\label{relax63}
\end{eqnarray}
We conclude that $\dot S\ge 0$ provided that $D,D'>0$. At equilibrium
$\dot S=0$ yielding ${\bf J}={\bf J}_{\xi}=0$. Equations
(\ref{relax55}) and (\ref{relax54}) imply
\begin{eqnarray}
\nabla\ln\rho+\frac{1}{2}\beta (\eta-\overline{\sigma})\nabla\biggl
(\frac{\overline{\sigma}}{y}\biggr )+\mu (\eta-\overline{\sigma})\nabla\xi=0,
\label{relax56}
\end{eqnarray}
\begin{eqnarray}
\beta\nabla\psi+\mu\nabla\overline{\sigma}=0.
\label{relax57}
\end{eqnarray}
The second equation is equivalent to
\begin{eqnarray}
\overline{\sigma}=-\frac{\beta}{\mu}\psi-\frac{\nu}{\mu}.
\label{relax58}
\end{eqnarray}
On the other hand, for any reference level $\eta_{0}$, Eq.
(\ref{relax56}) yields
\begin{eqnarray}
\nabla\ln\rho_{0}+\frac{1}{2}\beta (\eta_0-\overline{\sigma})\nabla\biggl
(\frac{\overline{\sigma}}{y}\biggr )+\mu (\eta_0-\overline{\sigma})\nabla\xi=0.
\label{relax59}
\end{eqnarray}
Subtracting Eqs.  (\ref{relax56})-(\ref{relax59}) and integrating, we obtain
\begin{eqnarray}
\ln\biggl (\frac{\rho}{\rho_{0}}\biggr )+\frac{1}{2}\beta
(\eta-\eta_0)\frac{\sigma}{y}+\mu (\eta-\eta_{0})\xi=A(\eta),
\label{relax510}
\end{eqnarray}
which can be written
\begin{equation}
\rho=\frac{1}{ Z({\bf r})}\chi(\eta)e^{-(\beta \frac{\overline{\sigma}}{
2y}+\alpha)\eta}e^{-\mu\xi\eta}.
\label{gibbs2}
\end{equation}
Thus, the stationary solution of the relaxation equations  is the
Gibbs state (\ref{rhoz}).

The relaxation equations are relatively complicated to solve, because
we need to solve $N$ coupled PDE, one for each level. Alternatively,
we can write down a hierarchy of equations for the moments of angular
momentum $\overline{\sigma^{n}}$. The first moment equations of the
hierarchy can be written
\begin{eqnarray}
\frac{\partial \overline{\sigma}}{\partial t}
+{\bf u}\cdot \nabla \overline{\sigma}
&=& \nabla\cdot \biggl\lbrace D\biggl\lbrack \nabla\overline{\sigma}+\frac{1}{
2}\beta \sigma_{2}\nabla\biggl (\frac{\overline{\sigma}}{y}\biggr
)+\mu\sigma_{2}\nabla\xi\biggr\rbrack\biggr \rbrace,\label{relax2bn}
\\
\frac{\partial\xi}{\partial t}
+{\bf u}\cdot \nabla {\xi}
&=&
\frac{\partial}{\partial z}\biggl (\frac{{\overline{\sigma}}^{2}}{4
y^{2}}\biggr ) +
\nabla \biggl\lbrack
D'(\beta\nabla\psi+\mu\nabla\overline{\sigma})\biggr\rbrack.
  \label{relax2b}
\end{eqnarray}
We are now led to a complicated closure problem because each equation
of the hierarchy involves the next order moments. For example, the
equation for $\overline{\sigma}$ involves $\sigma_{2}$ etc.  In the
two levels approximation, one has
$\sigma_{2}=(\overline{\sigma}-\lambda_{0})(\lambda_{1}-\overline{\sigma})$. On
the other hand, a Gaussian distribution of fluctuations at equilibrium
can be obtained by imposing that $\sigma_{2}$ is constant. In these
two particular cases, the equations (\ref{relax2bn}) are closed. More
generally, we must write down the higher moments of the hierarchy and
close them with a local maximum entropy principle as proposed by
Robert \& Rosier
\cite{Robert97}. If we implement this procedure up to second moment,
it leads to a Gaussian distribution. Its implementation to higher
moments is difficult. Furthermore, its physical justification is
unclear. In practice, we must come back to the $N$ coupled PDE for the
levels.

\subsection{Prior distribution and generalized entropy}
\label{SectionPrior}

In the statistical approach presented previously, we have assumed that
the system is rigorously described by the axisymmetric Euler equations
so that the conservation of all the Casimirs $I_{n}$ must be taken
into account. This corresponds to a freely evolving
situation. Alternatively, in the case of flows that are forced at
small-scales, Ellis {\it et al.} \cite{Ellis02} have proposed to
replace the conservation of all the Casimirs by the specification of a
prior distribution $\chi(\eta)$ encoding the small-scale forcing. This
approach has been further developed in Chavanis \cite{prior}. In this
approach, the constraints associated with the (fragile) moments
$I_{n>1}$ are treated canonically instead of microcanonically. By
contrast, the robust constraints (energy, circulation, helicity,
angular momentum) are still treated microcanonically. If we view the
levels $\eta$ of angular momentum as different species of particles,
this approach amounts to fixing the chemical potentials instead of the
total number of particles in each species. The idea is that the
ambient medium behaves as a reservoir of angular momentum: the
small-scale forcing and dissipation affect the conservation of the
moments of angular momentum $I_{n>1}$ while fixing instead the
canonical variables $\alpha_{n}$.

Therefore, in the present situation, the relevant entropy $S_{\chi}$
is obtained from the mixing entropy (\ref{entropym}) by making a
Legendre transform on the fragile moments $I_{n>1}$ \cite{prior}. If
we assume that the $\alpha_{n}$ in the variational principle
(\ref{vp}) are fixed by the ``reservoir'' (ambient medium), we can
define a relative entropy by
\begin{eqnarray}
S_{\chi}=S-\sum_{n>1}\alpha_{n}\overline{I_{n}}=S-\sum_{n>1}\alpha_{n}
\int \rho \eta^{n} \, dy dz d\eta.
\label{schi}
\end{eqnarray}
This is similar to the passage from the entropy $S$ (microcanonical
description) to the grand potential $\Omega=S-\mu N$ (grand
microcanonical description) in usual thermodynamics when the chemical
potential is fixed instead of the particle number. Using
Eq. (\ref{entropym}), we get
\begin{equation}
S_{\chi}=-\int \rho \biggl\lbrack
\ln\rho+\sum_{n>1}\alpha_{n}\eta^{n}\biggr\rbrack  \, dy dz d\eta.
\label{schi2}
\end{equation}
Introducing the function (\ref{chii}),
we obtain
\begin{equation}
S_{\chi}=-\int \rho \ln\biggl\lbrack
\frac{\rho}{\chi(\eta)}\biggr\rbrack \, dy dz d\eta.
\label{sa5}
\end{equation}
The function $\chi(\eta)$ is interpreted as a prior distribution of
angular momentum. It is a global distribution of angular momentum
fixed by the small-scale forcing. It must be regarded as {\it
given}. In this approach, the statistical equilibrium state is
obtained by maximizing the relative entropy (\ref{sa5}) while conserving
only the robust constraints. Thus, we write the variational problem as
\begin{eqnarray}
\delta S_{\chi}-\beta\delta E-\mu\delta
H-\gamma\delta\Gamma-\alpha \delta I
-\int \zeta(y,z)\delta\biggl (\int \rho d\eta\biggr ) dy dz=0.
\end{eqnarray}
We can now repeat the calculations of Sec. \ref{SectionGibbs} with
almost no modification. The only difference is that we regard the
$\alpha_{n}$'s as given. Therefore, the Gibbs state is determined by
Eq. (\ref{rhoz}) where $\chi(\eta)$ is fixed {\it a priori} by the
small-scale forcing. Recall that in the previous approach (freely
evolving flows), it had to be determined a posteriori from the initial
conditions (assumed known) by a complicated procedure. Here, the
specification of $\chi(\eta)$ automatically determines the function
$F$ by Eq. (\ref{var}) and then $C$ by Eq. (\ref{cf}). Thus, the
generalized entropy in $\overline{\sigma}$-space $S[\overline{\sigma}]=-\int
C(\overline{\sigma})
\, d{\bf r}$ is now determined by the small-scale forcing through the
prior $\chi(\eta)$ while in the preceding approach it was determined
by the initial conditions through the Casimirs $I_{n>1}$. Explicitly,
the generalized entropy is expressed as a function of $\chi$ by the
formula \cite{super}:
\begin{eqnarray}
C(\overline{\sigma})=-\int^{\overline{\sigma}}\lbrack
(\ln\hat{\chi})'\rbrack^{-1}(-x)dx.
\label{nf}
\end{eqnarray}
The equilibrium coarse-grained angular momentum $\overline{\sigma}$
maximizes the generalized entropy (\ref{entropy2})-(\ref{nf}) at fixed
robust constraints $E$, $\Gamma$, $H$ and $I$. In the present context,
the relaxation equations introduced in Sec. \ref{SectionAlgo} can
describe the relaxation of the coarse-grained flow towards statistical
equilibrium once the small-scale forcing has established a permanent
regime characterized by a prior distribution $\chi(\eta)$ determining
a generalized entropy $S[\overline{\sigma}]$. Since we are now
interested by the route to equilibrium we need to restore the
advective terms, so we write:
\begin{eqnarray}
\label{relax22b}
\frac{\partial \overline{\sigma}}{\partial t}
+{\bf u}\cdot \nabla \overline{\sigma}
&=&\nabla\cdot \biggl\lbrace D\biggl\lbrack
\nabla\overline{\sigma}+\frac{\beta}{C''(\overline{\sigma})}\nabla\biggl
(\frac{\overline{\sigma}}{2y}\biggr
)+\frac{\mu}{C''({\overline{\sigma}})} \nabla\xi\biggr\rbrack\biggr \rbrace,
\\ \nonumber
\frac{\partial\xi}{\partial t}
+{\bf u}\cdot \nabla {\xi}
&=&
\frac{\partial}{\partial z}\biggl (\frac{{\overline{\sigma}}^{2}}{4y^{2}}\biggr )+
\nabla\cdot  \biggl\lbrack
D'(\beta\nabla\psi+\mu\nabla\overline{\sigma})\biggr\rbrack.
\end{eqnarray}
The physical interpretation of these equations is quite different from
(\ref{relax22}). In Sec. \ref{SectionAlgo}, the relaxation equations
(\ref{relax22}) provide a {numerical algorithm} to determine any
nonlinearly dynamically stable stationary solution of the axisymmetric
equations, specified by the convex function $C$. In this context, only
the stationary solution for $t\rightarrow +\infty$ matters and the
evolution towards that state has no physical meaning (it is just the
engine of the algorithm). In the present section, the relaxation
equations (\ref{relax22b}) provide a description of the evolution of
the coarse-grained field, for all time $t$, in a medium where a
small-scale {forcing} imposes a prior distribution $\chi(\eta)$ (or a
generalized entropy $C(\overline{\sigma})$ in the reduced
$\overline{\sigma}$-space). These equations conserve only the robust
constraints and satisfy a generalized maximum entropy production
principle for the functional $S[\overline{\sigma}]$. Finally, the
relaxation equations of Sec. \ref{SectionMEPP} provide a description
of the evolution of the coarse-grained field, for all time $t$, of a
freely evolving system. These equations conserve all the constraints
(including the Casimirs) and satisfy a maximum entropy production
principle for the functional $S[\rho]$. Note that the relaxation
equations (\ref{relax22b}) can also be obtained from the moments
equations (\ref{relax2bn})-(\ref{relax2b}) of the ordinary statistical
theory by using the relation (\ref{sfw}) to express $\sigma_{2}$ as a
function of $\overline{\sigma}$. In a sense, this relation can be seen
as a closure relation imposed by a small-scale forcing. Thus, the
relaxation equations (\ref{relax22b}) are not simply numerical
algorithms; they can also provide a parameterization of axisymmetric
flows with a small-scale forcing. Their interest as numerical
algorithms (in the sense of Sec. \ref{SectionAlgo}) remains however
important in case of {\it incomplete relaxation} to construct stable
stationary solutions of the Euler equation which are not consistent
with the statistical theory (in cases where the evolution is
non-ergodic) in order to reproduce the observations, as discussed in
\cite{Chavanis03,prior,kupka,super}.

\section{Summary}

In this paper, we have developed new variational principles to study
the structure and the stability of equilibrium axisymmetric flows. We
have completely characterized the steady states of the inviscid
dynamics and found that there is an infinite number of solutions. We
have shown that each of these steady states extremizes a certain
functional and that maxima or minima of this functional correspond to
nonlinearly dynamically stable states. We have given analytical
solutions in some simple cases to illustrate our formalism. One of
these steady states (non-universal) will be reached on the
coarse-grained scale as a result of violent relaxation (chaotic
mixing). Our general approach must be contrasted from that of other
authors who obtained {\it particular} solutions of the Navier-Stokes
equation by means of phenomenological principles (minimum of enstrophy
for example in 2D turbulence, Beltramization for MHD and axisymmetric
flows,...). Such solutions are recovered as particular cases of our
formalism but many other solutions can emerge in practice depending on
the initial conditions, on the route to equilibrium (ergodicity) and
on the type of forcing. This is why we try to remain very general. Our
point is that there is no clear universality in 2D or axisymmetric
turbulence
\cite{Chavanis03,prior,kupka,super}.  In a second part, we have developed a
thermodynamical approach to determine the statistical equilibrium
states at some fixed coarse-grained scale. We found that the resulting
distribution can be divided in two parts: one universal part, coming
from the robust constraints, and one non-universal, which depends on
the initial conditions (Casimirs) for freely evolving systems or on a
prior distribution encoding non-ideal effects such as forcing and
dissipation. Finally, we have derived relaxation equations which can
be used either as numerical algorithm to compute stable stationary
solutions of the axisymmetric Euler equations, or to describe the
dynamics of the system (freely evolving or forced) at the
coarse-grained level. 

The main question regarding the application of our results to
realistic systems (such as the ones mentioned in the Introduction) is
the relevance of the use of the ideal (Euler) equation instead of the
true dissipative system. In fact, the presence of a small viscosity
does not preclude the applicability of our results. First of all,
since viscosity acts at small scales, its main effect is to erase the
fluctuations around the coarse-grained field.  Thus, it gives a
physical support for selecting the coarse-grained field which is at
large scales and which is relatively robust against viscosity. On the
other hand, we have shown that viscosity and coarse-graining act in a
similar manner so that they are not in opposition. In a very turbulent
flow, the diffusion acts only at small scales by dissipating
energy. By disregarding the details of the fine-grained dynamics, we
have a similar process where energy is lost in the small-scales but
accumulates in the large-scales. On the other hand, we have shown that
a (generalized) selective decay principle can be motivated either by
viscous effects or by coarse-graining. Indeed, a small viscosity or a
coarse-graining tend to increase the value of the $H$-functions
(fragile constraints) with only weak modification on the energy,
angular momentum, circulation and helicity (robust
constraints). Therefore, viscous effects do not break the nonlinear
dynamical stability results. On the contrary, they can precisely
explain (together with coarse-graining) {\it how} the system can reach
a maximum of an $H$-function at fixed robust constraints. Without
dissipation (viscosity or coarse-graining) this is not possible since
the Casimir functionals $S[\sigma]$ are rigorously conserved by the
Euler equation. We believe, however, that the main increase of the
fragile constraints (like enstrophy) is due to coarse-graining
\cite{JFM1} rather than molecular viscosity (in classical works on 2D
turbulence, it is argued instead that enstrophy is dissipated
essentially by viscosity). The main difference between viscous and
inviscid flows is that inviscid flows tend to a {\it strict}
stationary solution of the Euler equation (on the coarse-grained
scale) while, in the presence of a small viscosity, this large-scale
structure slowly diffuses and ultimately disapears. However, if
$\nu\rightarrow 0$, this happens on a long time scale that is not of
most physical interest. Note finally that forcing can act against
viscosity and maintain a steady state as for an inviscid evolution.

The other effect of viscosity, now regarding the statistical mechanics
approach, is to break the conservation of the Casimirs. This is a
problem for the original approach (Sec. \ref{SectionGibbs}) where it
is assumed that all the Casimirs are conserved. However, in the point
of view developed in Sec. \ref{SectionPrior}, we have replaced the
specification of the Casimirs by a prior distribution of angular
momentum. It corresponds to the non-universal part of the distribution
of fluctuations given by the Gibbs state (\ref{rhoz}). We have argued
that this prior is precisely determined by non-ideal effects such as
viscosity and forcing (in addition to the initial conditions and the
boundary conditions), i.e. by all the complicated features of
turbulence. Therefore, in this point of view, the existence of a
viscosity and a forcing can be taken into account phenomenologically
in the theory. On the other hand, it should be noted that the effect
of coarse-graining is similar to a turbulent viscosity. This is best
seen in the relaxation equations (\ref{relax22b}) which involve a
diffusion term with a ``turbulent viscosity'' $D$. However, our
approach shows that the adjunction of a turbulent viscosity to the
Euler equations in order to model turbulence is not sufficient as it
breaks the conservation of energy. Therefore, additional drift terms
arise in the relaxation equations to act against diffusion and lead to
a steady state \cite{prior}. There are other pieces of evidence for
the claim that the introduction of a coarse-graining procedure is
similar to a diffusive process. For example, recent numerical
simulations have shown that the Euler equation with a high wave-number
spectral truncation shows similar features as the Navier-Stokes
(dissipative) equation
\cite{Chichoune05}. This issue concerning the influence of viscosity
will be addressed more thoroughly in a second paper, where we confront
our prediction to experimental data, and use them to derive and
characterize the non-universal features of the equilibrium
distributions.

Finally, we will address the changes to be made to account for a
global rotation of the system. Taking the rotation vector to be
aligned in the $z$-direction, the Coriolis force will only add a term
$2 \Omega u$ on the left-hand-side of the first equation
(\ref{basic2}). Then, the conserved quantity will be
$\sigma'=rv+\Omega r^2$ instead of $\sigma = rv$. This is similar to
the use of a potential vorticity when doing the statistical mechanics
of two-dimensional rotating fluid instead of the usual vorticity
\cite{sw}. Similarly, the right-hand side of the second equation will
now be: $\partial_z[(v^2+ 2 \Omega v r)/r] = \partial_z[(v^2+ 2 \Omega
v r + \Omega^2 r^2)/r] = \partial_z[\sigma'^2/r^3]$. Consequently, all
the results in this paper will be valid provided that $\sigma'$ is
used instead of $\sigma$.

\section*{Acknowledgments}
We thank the programme national de Plan\'etologie, the GDR Turbulence
and the GDR Dynamo for support. We have benefited from numerous
discussion with our colleagues from GIT.

\appendix
\section{Derivation of conservation laws}
\label{AnnexeA}

In this Appendix, we prove the conservation laws used in the main
text. A cornerstone of the proof is the general identity:
\EQ
\int \chi\lbrace \phi,\psi\rbrace \, dy dz =-\int
\phi\lbrace\chi,\psi\rbrace \, dy dz \; ,
\label{id2}
\EN
which holds if one of the two fields $\chi$ or $\phi$ vanishes on the
boundary of the domain.
\begin{itemize}
\item Energy conservation: using the equations of motion and assuming that $\psi=0$ or $\xi=\sigma=0$ on the boundary of the domain, we have
\EQA
\label{energyvariation}
\dot E &=& \int  \psi\frac{\partial\xi}{\partial t} \, dy dz  + \int
\frac{\sigma}{2y}\frac{\partial\sigma}{\partial t} \,  dy dz ,\\
&=& \int \psi \left\lbrack - \lbrace\psi,\xi\rbrace -
\lbrace\frac{\sigma}{2y},\sigma\rbrace \right\rbrack -
\frac{\sigma}{2y} \lbrace\psi,\sigma\rbrace \, dydz \nonumber\\
&=& \int\left( - \xi\lbrace\psi,\psi\rbrace + \frac{\sigma}{2y}
\lbrace\psi,\sigma\rbrace - \frac{\sigma}{2y}
\lbrace\psi,\sigma\rbrace \right) \, dy dz \nonumber\\ \nonumber
&=&0,
\ENA
where we have used the identity (\ref{id2}) twice to obtain the third line.

\item Casimirs conservation: using the equations of motion and $\psi=0$ or $\sigma=0$ on the boundary, we will
show that all the moments of $\sigma$ are conserved:
\EQA
\label{casimirvariation}
\dot I_{n} &= & n \int \sigma^{n-1} \frac{\partial\sigma}{\partial t}
\, dy dz \, ,\\
&=& - n\int \sigma^{n-1}
\lbrace\psi,\sigma\rbrace \, dy dz \nonumber\\
&=& n\int \psi
\lbrace \sigma^{n-1},\sigma\rbrace \, dy dz \nonumber\\ \nonumber
&=&0,
\ENA
where we have used the
identity (\ref{id2}) in the third line.

\item Helicity conservation: using the equations of motion, we have
\EQA
\label{helicity}
\dot H&=&\int\biggl\lbrace F(\sigma)\frac{\partial\xi}{\partial t} +
\xi F'(\sigma)\frac{\partial\sigma}{\partial t}\biggr\rbrace \, dy dz
\, ,\\ \nonumber
&=&-\int  F(\sigma)\biggl\lbrack \lbrace\psi,\xi\rbrace+\lbrace
\frac{\sigma}{2y},\sigma\rbrace\biggr\rbrack-\int\xi F'(\sigma)
\lbrace\psi,\sigma\rbrace \, dy dz \, .
\ENA
Then,
\EQ
\int  F(\sigma)\lbrace \frac{\sigma}{2y},\sigma\rbrace \, dy dz =-
\int \frac{\sigma}{2y}\lbrace F(\sigma),\sigma\rbrace \, dy dz =0 \; ,
\label{helicitya}
\EN
if $\sigma$ or $F(\sigma)$ vanishes on the boundary of the domain. Therefore,
\EQA
\label{helicityb}
\dot H &=& -\int  F(\sigma)\lbrace\psi,\xi\rbrace dy dz -\int\xi F'(\sigma)
\lbrace\psi,\sigma\rbrace \, dy dz \\
&=& -\int  F(\sigma)\lbrace\psi,\xi\rbrace \, dy dz + \int\xi \lbrace
F(\sigma),\psi\rbrace \, dy dz \nonumber \\
&=& - \int \xi \lbrace F(\sigma),\psi\rbrace \, dy dz + \int\xi \lbrace
F(\sigma),\psi\rbrace \, dy dz \nonumber \\ \nonumber
&=& 0,
\ENA
where we have used identity (\ref{tr4}) in the second line and
identity (\ref{id2}) in the third line and assumed that $F(\sigma)=0$ or $\xi = 0$ on the boundary of the domain.

\end{itemize}

\section{Stability of solutions}
\label{VariationSecondes}

In section \ref{SectionDynstab}, we found that the functions
$\sigma_0(y,z)$ and $\xi_0(y,z)$ which extremize the functional
(\ref{entropy2}) are solutions of the following set of equations~:
\EQ
\label{SolutionStat}
\beta\psi_0 = - \mu\sigma_0 - {\nu} \qquad \mathrm{and} \qquad
-C'(\sigma_0) = {\beta}\frac{\sigma_0}{2y} + \mu\xi_0 + \alpha.
\EN
  However, only {\it maxima} of $S$ are nonlinearly 
dynamically stable. We need therefore to investigate the sign of the 
second order variations of $J=S-\beta E-\mu 
H-\nu\Gamma-\alpha I$.
Writing $\sigma=\sigma_0 + \delta \sigma$, $\xi =
\xi_0 + \delta \xi$ and $\psi=\psi_0 + \delta \psi$, one obtains for
all $\delta \sigma$ and $\delta \xi$:
\EQ
\delta^2 J[\sigma_0,\xi_0] = - \int dy dz \; \left[
\left(C''(\sigma_0)+\frac{\beta}{2y} \right) (\delta\sigma)^2 + 2 \mu
\delta\sigma \delta\xi + \beta \delta\psi \delta\xi\right] \; ,
\label{variationseconde}
\EN
with $\delta\xi = - \Delta_* \delta\psi$. Using the operators  $curl$
and ${\bf Curl}$ defined in \cite{MHD}, one can easily show that the
last term can be rewritten~:
\EQ
\beta \int \delta\psi \delta\xi \; dy dz = \beta \int \left[{\bf
Curl} \left(\frac{\delta\psi}{r}\right)\right]^2 \; dy dz \; .
\EN
Putting $\delta \sigma = 0$ in equation (\ref{variationseconde}), the
condition $\delta^2 J[\sigma_0,\xi_0] < 0$ thus implies that {\it 
$\beta$ must be positive}.  This is at variance with pure 2D 
hydrodynamics, where stable structures can exist at negative 
temperature and are the most relevant. Also, assuming $\delta \xi = 
0$, one finds that a maximum
of $J$ should satisfy the following condition:
\EQ
\int \; \left(C''(\sigma_0)+\frac{\beta}{2y} \right) (\delta\sigma)^2 \; dy dz > 0 \; ,
\EN
which is trivially fulfilled because $C$ is a convex function. One
cannot find a general condition on the value of $\mu$ in order for
$\delta^2 J[\sigma_0,\xi_0]$ to be negative but a sufficient condition
can be found by using the fact that  in (\ref{variationseconde}), the
last term in the integral is everywhere positive. Consequently, a
sufficient condition for $\delta^2 J < 0$ is:
\begin{center}
\EQA
&& \int dy dz \; \left[ \left(C''(\sigma_0)+\frac{\beta}{2y} \right)
(\delta\sigma)^2 + 2 \mu \delta\sigma \delta\xi \right] > 0 \\
\nonumber
\Leftrightarrow \int dy dz && \left[
\left(C''(\sigma_0)+\frac{\beta}{2y} \right) \left(\delta\sigma +
\frac{\mu}{C''(\sigma_0)+\frac{\beta}{2y}}\delta\xi\right)^2 -
\frac{\mu^2}{\left( C''(\sigma_0)+\frac{\beta}{2y}\right)^2}
(\delta\xi)^2 \right] > 0
\; ,
\ENA
\end{center}
for all $\delta \sigma$ and $\delta \xi$. This condition can
obviously be provided if $\mu =0$. A sufficient condition for
$\sigma_0$ and $\xi_0$, solution of (\ref{SolutionStat}), to be
maximum of $S$ is thus $\mu=0$ and $\beta >0$. However, this is only a very particular case.

\section{Stationarity of $\langle\sigma\rangle$ }
\label{sec_Z}

The most probable value of the distribution (\ref{rhoz}) can be written
\EQ
\langle \sigma \rangle = G(\Psi) = G[F^{-1}(\bar{\sigma})] = G\Bigl[F^{-1}\bigl(-\frac{\beta\psi + \gamma}{\mu}\bigr)\Bigr] \; ,
\EN
showing that $\langle \sigma \rangle = f(\psi)$ is a function of
$\psi$ alone. We now write the condition under which $\langle \sigma
\rangle$ is a stationary solution of the axisymmetric Euler equations. 
Comparison between Eqs. (\ref{tr10}) and (\ref{gibbsty}) shows that the
following relation must hold~:
\EQ
 - \frac{\beta}{\mu} \bar{\sigma}= \frac{1}{2} \frac{d}{d\psi}f^2 .
\EN
Using Eq. (\ref{gibbsre}), this can be rewritten:
\EQ
\label{fplusprob}
f(\psi) = \sqrt{\frac{\beta^2 \psi^2}{\mu^2} + 2 \frac{\gamma \beta}{\mu^2}\psi+c} \; ,
\EN
where $c$ is an integration constant. If we require that $\bar\sigma=\langle\sigma\rangle$ on the boundary of the domain  ($\psi=0$), then, using Eq. (\ref{gibbsre}), we get $f^2(0) = c = {\gamma^2}/{\mu^2}$. Substituting in 
Eq. (\ref{fplusprob}), we find that $G\circ F^{-1}$ is the identity so that $\langle\sigma\rangle=\bar\sigma$ (everywhere). This implies that $\langle\sigma\rangle$ is a stationary solution of the axisymmetric equations only when it coincides with  $\bar\sigma$ .

\section{Fluctuations of $\xi$}
\label{fluc}
\subsection{Generalities}

In this Appendix, we try to develop the statistical mechanics approach
in the general case, without ignoring the fluctuations of $\xi$. Since
$\xi$ is not conserved by the axisymmetric equations ($D\xi/Dt\neq
0$), this may invalidate the use of a statistical theory to predict
its fluctuations, so that our approach is essentially phenomenological
and explanatory.  We introduce $\rho({\bf r},\eta,\nu)$, the density
probability of finding the values $\sigma=\eta$ and $\xi=\nu$ in ${\bf
r}$ at equilibrium. Then, the coarse-grained fields are
$\overline{\sigma}=\int \rho\eta \, d\eta d\nu$,
$\overline{\xi}=\int\rho\nu \, d\eta d\nu$ and the local normalization
is $\int \rho \, d\eta d\nu=1$. We introduce the mixing entropy
\begin{equation}
S[\rho]=-\int \rho\ln\rho\ dy dz d\eta d\nu.
\label{entropygene}
\end{equation}
  As usual, the fluctuations of $\psi$ will be neglected because it is 
an integrated quantity of the primitive field $\xi$. The integral 
constraints can be re-expressed as
\EQA
\overline{E} &=& \frac{1}{2} \int \rho \nu \psi \; dy dz d\eta d\nu \, + \,
\frac{1}{ 4}\int \rho
\frac{\eta^2}{y} \; dy dz d\eta d\nu \; ,
\label{energy3}
\\
& & \overline{H_{n}} = \int\rho\nu \eta^n \; dy dz d\eta d\nu \; ,
\label{helicity3}
\\
\overline{I_{n}} &=& \int \rho \eta^{n} \; dy dz d\eta d\nu,\; ,
\qquad \overline{\Gamma} =\int \rho\nu \; dy dz d\eta d\nu.
\;  
\ENA
The most probable distribution at metaequilibrium is therefore
obtained by maximizing the entropy at fixed $E$, $H_n$ and
$I_{n}$. We introduce Lagrange multipliers and write the variational
principle in the form
\EQ
\delta S-\beta\delta\overline{E}-\sum_{n}\mu_n\delta
\overline{H_n}-\sum_{n}\alpha_n \delta \overline{I_n} - \int
\zeta(y,z)\delta\biggl (\int \rho \; d\eta d\nu \biggr )\; dy dz =0.
\EN
The variations on $\rho$ yield the Gibbs state
\begin{equation}
\rho=\frac{1}{Z(y,z)}e^{-\beta [\nu\psi+\frac{\eta^2}{4y}]-\alpha G(\eta)-\mu F(\eta)\nu},
\label{gibbs}
\end{equation}
where $\mu F(\eta)=\sum_{n}\mu_{n}\eta^{n}$ and $\alpha
G(\eta)=\sum_{n}\alpha_n\eta^n$  . The ``partition function'' is
determined by the local normalization condition yielding
\begin{equation}
Z=\int e^{-\beta[\nu\psi+\frac{\eta^2}{4y}]-\alpha G(\eta)-\mu F(\eta)\nu}d\eta d\nu.
\label{partition5}
\end{equation}
The coarse-grained fields $\bar\sigma$ and $\bar\xi$ are the {\it
averaged values} of the distribution (\ref{gibbs}).  This approach
predicts that the distribution of fluctuations of pseudo-vorticity is
exponential $\sim e^{\Lambda(y,z,\eta)\nu}$ so that, depending on the
sign of $\Lambda$, it diverges either for $\nu\rightarrow +\infty$ or
$\nu\rightarrow -\infty$. The problem of the smoothness of the
vorticity is still an unresolved issue for the Navier-Stokes equation
related to the existence and uniqueness of its solutions. In two
dimensions, it can be shown that it is bounded \cite{Doering} whereas
in three dimensions, very little is known. In our case, which is
intermediate between these two cases, we can infer that the vorticity
will be bounded as we have seen that in the axisymmetric case, the
vorticity has to vanish in the long time limit (see section
\ref{SectionSelective}). The range of integration for the variable
$\nu$ is thus restricted to finite values: $|\nu| < \lambda$.

We now look for {\it extremum values} of the distribution
(\ref{gibbs}).To study this problem, we write:
\EQA
\rho(y,z,\eta,\nu) &=& \frac{1}{Z} \exp[-\cal{F}(\eta,\nu)] \; , \\ \nonumber
\text{with} \qquad  \cal{F} &=& \nu [\beta \psi + \mu F(\eta)] + \frac{\beta \eta^2}{4y} + \alpha G(\eta) \; .
\ENA
We start to search for extrema of the distribution in the interior of
the domain $] -\xi_{m},\xi_{m}[ $ and call them
$\langle\sigma\rangle,\langle\xi\rangle$. One can check that they obey
the equations:
\EQA
\label{ext}
\frac{\partial {\cal F}}{\partial \eta} \Big\vert_{\langle\sigma\rangle, \langle\xi\rangle} &=& \frac{\beta \langle\sigma\rangle}{2y} + \alpha G\,'(\langle\sigma\rangle) + \mu \langle\xi\rangle F\,'(\langle\sigma\rangle) = 0,\\ \nonumber
\frac{\partial {\cal F}}{\partial \nu} \Big\vert_{\langle\sigma\rangle, \langle\xi\rangle} &=& \beta \psi + \mu F(\langle\sigma\rangle) = 0.
\ENA
These extremum fields $\langle\sigma\rangle$
and $\langle\xi\rangle$ are stationary states of the Euler
axisymmetric equation (with families indexed through the conservation
laws) while the averaged states $\bar{\sigma}$ and $\bar{\xi}$ are not
in general. The stability of these extremum states can be found in
principle by considering the second variations of ${\cal F}$. Here, we
prefer to use the following trick. We introduce the functions :
\EQA
\label{trick1}
Q(\sigma_*,\xi_*) &=& \beta\psi+\mu F(\sigma_*) \; , \\ \nonumber
R(\sigma_*,\xi_*) &=& \frac{\beta\sigma_*}{2y} + \alpha G'(\sigma_*)+\mu \xi_* F'(\sigma_*).
\ENA
Then, for any $\sigma_*$ and $\xi_*$, one has :
\EQA
\label{trick2}
{\cal F}(\eta,\nu) - {\cal F}(\sigma_*,\xi_*) &=& - Q(\sigma_*,\xi_*) (\nu-\xi_*)-R(\sigma_*,\xi_*)(\eta-\sigma_*) - \frac{\beta}{4y}(\eta-\sigma_*)^2 - \mu(\nu-\xi_*)(\eta-\sigma_*)F'(\sigma_*) \\ \nonumber &-& \alpha\left[G(\eta)-G(\sigma_*)-(\eta-\sigma_*)G'(\sigma_*)\right]-\mu\nu\left[F(\eta)-F(\sigma_*)-(\eta-\sigma_*)F'(\sigma_*)\right].
\ENA
Choosing $\sigma_*=\langle\sigma\rangle$ and
$\xi_*=\langle\xi\rangle$, we have $Q=R=0$, so that the probability
function simply becomes:
\EQA
\label{trick3}
\rho = \frac{1}{Z_*(y,z)} \exp\Bigl[&-&\frac{\beta}{2y} (\eta-\langle\sigma\rangle)^2 - \mu(\nu-\langle\xi\rangle)(\eta-\langle\sigma\rangle) F'(\langle\sigma\rangle) - \alpha\left[G(\eta) - G(\langle\sigma\rangle)
- (\eta-\langle\sigma\rangle) G'(\langle\sigma\rangle) \right] \\ \nonumber
&-& \mu\nu\left[F(\eta) - F(\langle\sigma\rangle) - (\eta-\langle\sigma\rangle) F'(\langle\sigma\rangle)\right] \Bigr] , \\ \nonumber
Z_* &=& Z e^{{\cal F}(\langle\sigma\rangle,\langle\xi\rangle)}.
\ENA
Due to the $e^{\Lambda\nu}$ dependence of the density probability, one
can check that the extremal states of the Gibbs probability
distribution are saddle points (stable in one direction, unstable in
the other), except when $F$ is constant and $\beta\psi+\mu F=0$
(leading to $\bar\xi=0$), in which case they are stable
states for positive temperature $\beta>0$. The fields $\langle \sigma
\rangle$ and $\langle \xi
\rangle$ thus are not real extrema of the distribution.  Moreover,
one sees that when $F$ is constant with $\beta\psi+\mu F=0$ and $G$ is
linear, the probability distribution of $\eta$ is a Gaussian in the
variable $(\eta-\langle\sigma\rangle)^2$ and the probability
distribution of $\nu$ is uniform. Therefore, the most probable state $\langle
\sigma\rangle$ coincides with the mean state $\bar\sigma$. However, this is not the generic
case.

We now look at possible extrema on the frontier of the domain of
integration, for $\nu = \pm \lambda$. It is obvious that if it exists a
physical bound on the vorticity, it must depend on the shape of
velocity field: $\lambda = \lambda(\bar{\sigma},\psi)$. Assuming this function
to be known, we can write the conditions that must satisfy an extremum
$(\sigma_0,\xi_0)$ of $\rho$ located on the frontier of the
integration domain:
\EQA
\frac{\partial \cal{F}}{\partial \eta} \Big\vert_{\sigma_0,\xi_0} &=& \frac{\beta \sigma_0}{2y} + \alpha G\,'(\sigma_0) + \mu \xi_0 F\,'(\sigma_0) = 0, \\ \nonumber
\xi_0 &=& \pm \lambda (\bar{\sigma},\psi).
\ENA
We note that these fields are not stationary states of the Euler
axisymmetric equation, contrary to $\langle\sigma\rangle$ and $\langle
\xi\rangle$. To decide which of the two couples $(\langle \sigma
\rangle,\langle \xi \rangle)$ or $(\sigma_0,\xi_0)$ is the most
probable state of the distribution $\rho$, one has to compare the value of the function ${\cal F}$ at
these two points:
\EQ
{\cal F}(\sigma_0,\xi_0) = \xi_0[\beta \psi + \mu \{F(\sigma_0)-\sigma_0 F'(\sigma_0)\}] + \alpha [G(\sigma_0) - \sigma_0 G'(\sigma_0)]
\EN
\EQ
\text{and} \qquad {\cal F}(\langle \sigma \rangle,\langle \xi \rangle) = \alpha[G(\langle \sigma \rangle) - \langle \sigma \rangle G'(\langle \sigma \rangle)] - \mu F'(\langle \sigma \rangle) \langle \sigma \rangle \langle \xi \rangle.
\EN
From these expressions, it is not possible to decide which one of these two values is the smallest (corresponding to a maximum value for $\rho$) in the general case. However, in the special case where $G\propto \sigma$ and  $F\propto \sigma$, we have
\EQ
{\cal F}(\sigma_0,\xi_0) = \xi_0\beta \psi,
\EN
\EQ
\text{and} \qquad {\cal F}(\langle \sigma \rangle,\langle \xi \rangle) = -\mu\langle\sigma\rangle\langle\xi\rangle=\langle\xi\rangle\beta \psi,
\EN
where we have used Eq. (\ref{ext})-b to obtain the last equality.  Since
$|\langle\xi\rangle|\le \lambda$, one obtains the most probable state
on the boundary of the domain of integration.  More generally, since
($\langle\sigma\rangle$,$\langle
\xi\rangle$) corresponds to a saddle point of ${\cal F}$, the relevant solution to consider should be the solution ($\sigma_{0}$, $\xi_{0}$) 
where $\xi$ reaches its maximum bound. Therefore, this approach
suggests that the equilibrium states of axisymmetric flows are those
that maximize $\xi$ (the toroidal component of the vorticity). Since
the dissipation of kinetic energy is equal to the space integral of
the squared vorticity (see Sec. \ref{SectionSelective}), our conclusion
resembles the assumption made by Malkus \cite{Malkus}, followed by
Howard \cite{Howard} and Busse
\cite{Busse}, who calculated bounds on the kinetic energy dissipation
for thermal convection problems by maximizing the dissipation on a
manifold that includes the solutions of the problem. This principle of
maximal dissipation has been extended to a purely chaotic system by
\cite{Petrelis04} and in this case too, the observed equilibrium
solutions are very close to that calculated by a maximization of the
dissipation. It is however interesting to notice that in these
approaches, the maximum dissipation is {\it assumed} and the shape of the
equilibrium field is derived while in our approach we show
that the equilibrium state (which maximizes the entropy) is  the one with
a maximal vorticity field. The drawback of our approach, however, is that we
do not have an explicit form for the equilibrium solution, unless we
know how to derive the bound on the vorticity.

\subsection{Examples}

A few examples can be given to illustrate the points developed above.
For simplicity, let us consider first the case with $F=1$, where
$\eta$ and $\nu$ become independent. In such case, the probability
distribution function is :

\EQA
\rho=\frac{1}{Z}e^{-\nu(\beta\psi+\mu)-\frac{\beta\eta^{2}}{4 y}-\alpha G(\eta)},
\ENA
and the partition function factorizes into $Z=Z_{\nu}Z_{\eta}$ with :
\EQA
Z_{\nu}=\int_{-\lambda}^{+\lambda} 
e^{-\nu(\beta\psi+\mu)}d\nu=\frac{2}{\beta\psi+\mu}\sinh\lbrack
\lambda (\beta\psi+\mu)\rbrack \; .
\ENA
Following the discussion of the previous section, we have introduced a
symmetrical cut-off $\lambda$. This situation is similar to the
Turkington model in 2D turbulence, see \cite{Chavanis03}. By
integration, one finds 
\EQA
\overline{\xi}=\frac{1}{Z_{\nu}}\int_{-\lambda}^{+\lambda} \nu
e^{-\nu(\beta\psi+\mu)}d\nu=-\frac{\partial\ln Z_{\nu}}{\partial
(\mu+\beta\psi)}=\lambda L\lbrack -\lambda(\beta\psi+\mu)\rbrack \; ,
\ENA
where
\EQA
L(x)=\tanh^{-1}(x)-\frac{1}{x} \; ,
\ENA
is the Langevin function. For the other part of $Z$ we get :
\EQA
Z_{\eta}=\int_{-\infty}^{+\infty}e^{-\frac{\beta\eta^{2}}{4 y}-\alpha
G(\eta)}d\eta \; .
\ENA

\par $\bullet$ Case $G(\eta)=\eta$, $F=1$: in this case, the extremal state is $\psi=-\mu/\beta$, $\langle\sigma\rangle=-2\alpha y/\beta$ or
$\sigma_{0}=-2\alpha y/\beta$, $\xi_{0}=\pm\lambda$. To derive
the mean state, we first compute :
\EQA
Z_{\eta}=2\biggl (\frac{\pi y}{\beta}\biggr )^{1/2}e^{\frac{\alpha^{2}}{\beta}y} \; .
\ENA
The mean state may then be found from :
\EQA
\overline{\sigma}=\frac{1}{Z_{\eta}}\int_{-\infty}^{+\infty}\eta
e^{-\frac{\beta\eta^{2}}{4 y}-\alpha \eta}d\eta =-\frac{\partial\ln
Z_{\eta}}{\partial \alpha}=-\frac{2\alpha}{\beta}y \; .
\ENA
The mean state $\bar\sigma$ therefore coincides with the extremal state
$\langle\sigma\rangle$ if $\psi=-\mu/\beta$ or $\sigma_{0}$. The mean state $\bar\xi$ is equal to zero if $\psi=-\mu/\beta$ and is lower than $|\xi_{0}|=\lambda$ (in absolute value) otherwise.

\par $\bullet$ Case  $G(\eta)=\eta+k\eta^{2}$, $F=1$: in this case, the extremal state is $\psi=-\mu/\beta$, $\langle\sigma\rangle=-2\alpha y/(\beta+4k\alpha y)$ or
$\sigma_{0}=-2\alpha y/(\beta+4k\alpha y)$, $\xi_{0}=\pm\lambda$.
For the partition function, we have :
\EQA
Z_{\eta}=2\biggl (\frac{\pi y}{\beta+4\alpha k y}\biggr
)^{1/2}e^{\frac{\alpha^{2}y}{\beta+4\alpha k y}} \; ,
\ENA
and
\EQA
\overline{\sigma}=-\frac{\partial\ln Z_{\eta}}{\partial \alpha}=\frac{2\beta
y(\alpha-k)+4\alpha k y^{2}(\alpha-2k)}{ (\beta +4\alpha k y)^{2}} \; .
\ENA
Therefore, the mean state does not coincide with the extremal state.

\par $\bullet$ Case $F(\eta)=G(\eta)=\eta$: in this case, the extremal state is $\psi=-(\mu/\beta) \langle\sigma\rangle $,
$\langle\xi\rangle =-\alpha/\mu-(\beta/2y\mu)\langle\sigma\rangle$ or
$\xi_{0}=-\alpha/\mu-(\beta/2y\mu)\sigma_{0}$, $\xi_{0}=\pm\lambda$.
Integrating the partition function first with respect to $\eta$, we
get :
\EQ
Z=\sqrt{\frac{4\pi y}{\beta}}\int_{-\lambda}^{+\lambda} d\nu
\exp\Bigl[-\nu \beta \psi + \frac{(\alpha+\mu\nu)^2y}{\beta}\Bigr] \; .
\EN
Using this, we find the mean state as :
\EQA
\bar{\xi} &=& \frac{1}{Z} \sqrt{\frac{4\pi y}{\beta}} 
\int_{-\lambda}^{+\lambda} d\nu \nu \exp\Bigl[-\nu \beta \psi + 
\frac{(\alpha+\mu\nu)^2y}{\beta}\Bigr] \\ \nonumber
&=& \frac{1}{Z}\sqrt{\frac{4\pi y}{\beta}} \frac{\beta}{2\mu^2 y} 
\Bigl\{ \int_{-\lambda}^{+\lambda} d\nu [-\beta\psi + \frac{2\mu 
y}{\beta}(\alpha+\mu \nu)] \exp\Bigl[-\nu \beta \psi + 
\frac{(\alpha+\mu\nu)^2y}{\beta}\Bigr] \Bigr\} + \frac{\beta^2 
\psi}{2\mu^2 y} - \frac{\alpha}{\mu}	\\ \nonumber
&=& \frac{1}{Z}\sqrt{\frac{4\pi y}{\beta}} \frac{\beta}{\mu^2 y} 
\exp\Bigl[\frac{(\alpha^2+\mu^2\lambda^2)y}{\beta}\Bigr] 
\sinh\Bigl[\lambda \beta \psi - \frac{2 \alpha \mu \lambda 
y}{\beta}\Bigr]+ \frac{\beta^2 \psi}{2\mu^2 y} - \frac{\alpha}{\mu} \; ,
\ENA
and
\EQ
\bar{\sigma}=-\frac{\partial \ln Z}{\partial \alpha} =
-\frac{2y}{\beta} (\alpha+\mu \bar{\xi}) \; .
\EN
When the cut-off $\lambda$ is taken into account, the mean state does
not coincide with the extremal state.

\bibliographystyle{apsrev}
\bibliography{bibliographie4}

\end{document}